\RequirePackage{lineno}
\documentclass[a4,nofootinbib,notitlepage,superscriptaddress]{revtex4-1}
\usepackage{bm}
\usepackage{lineno}
\usepackage[pdftex]{graphicx}

\newcommand{\I}{\mathrm{i}}

\usepackage{amsmath,amsfonts,amssymb,mathtools}
\usepackage{amsthm}

\usepackage{fancyhdr}
\usepackage{float}
\usepackage{mathrsfs}
\usepackage{setspace}

\usepackage[normalem]{ulem}

\usepackage{physics}
\usepackage{xcolor}

\usepackage[utf8]{inputenc}
\usepackage{pgfplots}
\usepgfplotslibrary{groupplots}
\usepgfplotslibrary{fillbetween}
\usepackage{apptools}
\AtAppendix{\counterwithin{theorem}{section}}

\DeclarePairedDelimiterXPP{\sfTr}[1]{\mathsf{Tr}}{[}{]}{}{#1}
\DeclarePairedDelimiterXPP{\sfTrAbs}[1]{\mathsf{TrAbs}}{[}{]}{}{#1}
\DeclarePairedDelimiterXPP{\opTr}[1]{\mathrm{Tr}}{[}{]}{}{#1}
\DeclarePairedDelimiterXPP{\bbTr}[1]{\mathbb{T}\mathrm{r}}{[}{]}{}{#1}

\newtheorem{theorem}{Theorem}
\newtheorem{lemma}[theorem]{Lemma}

\newtheorem{corollary}[theorem]{Corollary}

\def\sfW{\mathsf{W}}
\def\bbS{\mathbb{S}}
\def\bbL{\mathbb{L}}
\def\bbW{\mathbb{W}}
\def\Lra{\Leftrightarrow}

\def\ANU{Centre for Quantum Computation and Communication Technology, Department of Quantum Science, Australian National University, Canberra, ACT 2601, Australia.}
\def\NTU{School of Physical and Mathematical Sciences, Nanyang Technological University, Singapore 639673, Republic of Singapore}
\def\UEC{Graduate School of Informatics and Engineering, The
  University of Electro-Communications, Tokyo 182-8585, Japan}

\begin{document}
\title{Efficient computation of the Nagaoka--Hayashi bound for
  multi-parameter estimation with separable measurements}

\author{Lorc\'{a}n O. Conlon}
\email{lorcan.conlon@anu.edu.au}
\affiliation{\ANU}
\author{Jun Suzuki}
\email{junsuzuki@uec.ac.jp}
\affiliation{\UEC}
\author{Ping Koy Lam}
\affiliation{\ANU}
\affiliation{\NTU}
\author{Syed M. Assad}
\email{cqtsma@gmail.com}
\affiliation{\ANU}
\affiliation{\NTU}

\date{\today}
\begin{abstract}
  Finding the optimal attainable precisions in quantum multiparameter
  metrology is a non-trivial problem. One approach to tackling this
  problem involves the computation of bounds which impose limits on
  how accurately we can estimate certain physical quantities. One such
  bound is the Holevo Cram\'er--Rao bound on the trace of the mean
  squared error matrix. The Holevo bound is
  an asymptotically achievable bound when one allows for any
  measurement strategy, including collective measurements on many
  copies of the probe. In this work we introduce a tighter bound for estimating multiple parameters simultaneously when performing separable measurements on finite copies of the probe. This makes it more relevant in terms of experimental accessibility. We show that this bound can be efficiently computed by casting it as a semidefinite program. We illustrate our bound with several examples of
  collective measurements on finite copies of the probe. These
  results have implications for the necessary requirements to saturate the
  Holevo bound.
\end{abstract}

\maketitle
\section{Introduction}
Quantum mechanics simultaneously offers unique opportunities and
limitations for metrology. On the one hand, uniquely quantum
mechanical effects such as squeezing allow greater measurement
sensitivity than is classically possible~\cite{giovannetti2004quantum,
  giovannetti2011advances}. This is most evident in the search for
gravitational waves, where the injection of squeezed light into LIGO
has resulted in a significant increase in
sensitivity~\cite{aasi2013enhanced}. Quantum resources have been shown
to offer enhanced measurement capabilities in a range of applications,
including optical interferometry~\cite{caves1981quantum,
  barnett2003ultimate, dorner2009optimal, kacprowicz2010experimental,
  demkowicz2009quantum}, quantum superresolution
\cite{tsang2016quantum, tsang2019resolving}, quantum-enhanced phase
tracking~\cite{yonezawa2012quantum, zhang2019quantum} and quantum
positioning~\cite{giovannetti2001quantum, lamine2008quantum} to name
but a few examples. Fundamental limits to single-parameter measurement
precisions can be computed using the quantum version of the
Cram\'er--Rao bounds~\cite{helstrom1968minimum, helstrom1967minimum,
  belavkin1976generalized}. On the other hand the uncertainty
principle places fundamental limits on how well two or more non-commuting
observables can be simultaneously
measured~\cite{robertson1929uncertainty}. Many of the applications
of quantum estimation require the simultaneous measurement of multiple
parameters~\cite{cimini2019quantum,vrehavcek2017multiparameter}, which in general will not commute with each other. This
means that a measurement that is optimal for one parameter may not be
optimal for another. This places a limit on the precision with which
we can measure them simultaneously ~\cite{ragy2016compatibility,
  szczykulska2016multi,kull2020uncertainty,demkowicz2020multi,suzuki2020quantum}. Thus
in an effort to fully exploit quantum resources in real-world
applications, there has been great
experimental~\cite{steinlechner2013quantum,hou2016achieving,roccia2017entangling,liu2018loss,hou2018deterministic}
and theoretical interest in quantum multi-parameter
estimation~\cite{humphreys2013quantum,Genoni2013,Crowley2014,
  gagatsos2016gaussian, baumgratz2016quantum, chrostowski2017super,
  pezze2017optimal,suzuki2016explicit,
  szczykulska2017reaching,albarelli2020perspective,assad2020accessible,
  tsang2019quantum, carollo2019quantumness}. Reviews of recent
progress on the subject are given in Refs.~\cite{sidhu2020geometric,
  polino2020photonic,demkowicz2020multi,suzuki2020quantum,albarelli2020perspective}.

Except for special cases involving qubits~\cite{suzuki2015parameter}
or estimating Gaussian amplitudes~\cite{bradshaw2018ultimate,
  bradshaw2017tight}, in general the problem of finding the optimal
measurement that minimises the sum of the mean squared error in
multi-parameter estimation is a non-trivial problem. Instead, one
resorts to finding bounds on these
errors~\cite{hayashi2005asymptotic}. Some of these bounds are the
bounds based on the symmetric logarithmic derivatives
(SLD)~\cite{helstrom1968minimum, helstrom1967minimum} and the right
logarithmic derivatives~\cite{yuen1973} as well as the
Gill--Massar~\cite{gill2005state} bound. While these bounds are easy
to compute, they are in general not tight. A tighter bound for the sum
of the mean squared error which can be achieved in the asymptotic
limit is given by the Holevo Cram\'er--Rao
bound~\cite{holevo2011probabilistic}. The computation of the Holevo
bound was recently cast as a semidefinite program which has made it
easy to compute. This was first performed for the Gaussian amplitude
estimation problem~\cite{bradshaw2018ultimate} and was later
generalised to an arbitrary model~\cite{Albarelli2019}. Furthermore
analytic expressions which upper and lower bound the Holevo bound have
recently been found~\cite{sidhu2021tight}. In some special cases the
measurement strategy required to reach the Holevo bound is known, for
example with pure state probes~\cite{Matsumoto2002} or for estimating
a single parameter.

In general the Holevo bound is only asymptotically
achievable~\cite{kahn2009local, yamagata2013quantum,
  yang2019attaining}, requiring a {\em collective measurement} over
infinitely many copies of the probe state. A collective measurement
here means that all copies of the probe state are measured
simultaneously. In contrast a {\em separable measurement} restricts the
probe states to be measured individually. In practice collective
measurements are extremely challenging to perform and are not
accessible to most experimental teams. Thus it would be useful to have
a tighter bound on the minimum achievable error when restricted to
separable, single-copy measurements. One such bound for simultaneously
estimating two-parameters was introduced
by Nagaoka~\cite{nagaoka2005new}. This bound is at least as tight as the
Holevo bound and it can be saturated for probes in a two-dimensional
Hilbert space~\cite{nagaoka2005generalization}. However, just like the
Holevo bound, Nagaoka's bound is not an explicit bound---it requires a
further non-trivial minimisation.

In this work we generalise the Nagaoka bound to estimating more than
two parameters, and we call this generalised bound the Nagaoka--Hayashi
bound. This bound applies to separable measurements on a finite number
of copies of the probe state, unlike the Holevo bound which, as
mentioned above, is only asymptotically attainable in general. We
further show that the minimisation required in the Nagaoka--Hayashi
bound can computed using a semidefinite program. This makes its
computation accessible. We illustrate our results with two examples
which highlight some of the interesting features of finite copy metrology which are
inaccessible with conventional techniques. In both of these
examples we are able to find the positive operator valued measure
(POVM) which saturates the bound, however whether this is always
possible remains an open question.

\section{Results}
Consider an $n$-parameter family of states
$\left\{S_\theta | \theta\in\Theta\subseteq\mathbb{R}^n  \right\}$ in
a finite $d$ dimensional Hilbert space $\mathcal{H}_q$ with
$\theta=\left( \theta_1,\ldots, \theta_n \right)^\intercal$ denoting
the $n$ independent \emph{true values} that we wish to estimate. Let
$\Pi=(\Pi_1,\ldots, \Pi_M)^\intercal$ be a column vector of $M$ POVM
elements, where $(\cdot)^\intercal$ denotes partial transpose with respect to the classical subsystem. The quantum
  operators $\Pi_m$ are not transposed. This means $\Pi_m \geq 0$ and
$\sum_m \Pi_m=1$. Each outcome $m$ assigns an \emph{estimated value}
for $\theta_j$ through the classical estimator function
$\hat{\theta}_{jm}$. The standard measure of estimation error when
restricted to separable measurements is through the $n$-by-$n$ mean
squared error (MSE) matrix $\mathsf{V}_\theta(\Pi,\hat\theta)$ with entries
\begin{align}
  \label{eq:parMSE}
  \left[\mathsf{V}_\theta(\Pi,\hat\theta)\right]_{jk}= \sum_{m}
  \left(\hat{\theta}_{jm}-\theta_j \right)
  \left(\hat{\theta}_{km}-\theta_k \right)\opTr{S_\theta
  \Pi_m}\;,\qquad \text{for } j,k=1,\dots,n\;.
\end{align}
The notation $\opTr{\cdot}$ in serif font is used to represent the trace of an operator in
 $\mathcal{H}_{q}$, the Hilbert space of the quantum system. 
For brevity of notation, hereafter we drop the argument and write the MSE matrix as
$\mathsf{V}_\theta$. We aim to minimise the trace of the MSE matrix under the
condition that our estimates are locally unbiased
\begin{align}
  \label{eq:Piunbiased}
  \sum_m \opTr{S_\theta \Pi_m} \hat{\theta}_{jm}=\theta_j\qquad\text{and}\qquad
 \sum_m  \frac{\partial}{\partial\theta_k} \opTr{S_\theta \Pi_m} \hat{\theta}_{jm}=\delta_{jk}\;.
\end{align}
The Nagaoka bound for two-parameter estimation
gives a lower bound on the trace of the MSE matrix
as~\cite{nagaoka2005new}
\begin{align}
\label{eq_hol2N}
\sfTr{\mathsf{V}_\theta} \geq \min_{X} \big\{ \opTr{S_\theta X_1 X_1+S_\theta X_2 X_2} +
  \text{TrAbs}\,S_\theta[X_1,X_2] \big\}  \eqqcolon c_{\text{N}}\;,
\end{align}
where the sans-serif font $\sfTr{\cdot}$ denotes the trace of a classical matrix in $\mathcal{H}_{c}$, $\text{TrAbs}\,A$ is the sum of the absolute values of the
eigenvalues of the operator $A$, and $X=(X_1,X_2,...,X_n)^\intercal$
is a vector of Hermitian estimator observables $X_j$ that satisfy the
locally unbiased condition at $\theta$
\begin{align}
  \label{eq:Xunbiased}
  \opTr{S_\theta X_j}= \theta_j \qquad\text{and}\qquad \frac{\partial}{\partial \theta_j} \opTr{S_\theta X_k}=\delta_{jk}\;.
\end{align}
The Nagaoka bound was conjectured to be a tight bound for
$\sfTr{\mathsf{V}_\theta}$~\cite{nagaoka2005generalization}. 
\subsection{Computable multi-parameter bound}
\label{sec:mainres}
As we shall shortly prove, the Nagaoka bound can be
generalised to more than two parameters. This result is stated as the
following theorem.
\begin{theorem}[Nagaoka--Hayashi bound]
  \label{thm:cnh}
  Let $\mathsf{V}_\theta$ be the MSE matrix of an unbiased estimate of
  $\theta$ for a separable measurement on a model $S_\theta$. Then
  the trace of $\mathsf{V}_\theta$ is bounded by
  \begin{equation}
  \begin{split}
\sfTr{\mathsf{V}_\theta}\geq & \min_{\mathbb{L},\,X}\left\{
                          \bbTr{\mathbb{S}_\theta
                          \mathbb{L}}\,\big|\, \mathbb{L}_{jk}=\mathbb{L}_{kj}\, \mathrm{
       Hermitian, }\,\mathbb{L}\geq {X} X^\intercal,\, X_j
    \,\mathrm{Hermitian\,satisfying\,\eqref{eq:Xunbiased} }
    \right\} \eqqcolon c_\mathrm{NH}\;,\label{eq:gnb}
\end{split}
\end{equation}
where $\mathbb{S}_\theta= {1}_n\otimes S_\theta$ and $\mathbb{L}$
is an $n$-by-$n$ matrix of Hermitian operators $\mathbb{L}_{jk}$.
\end{theorem}
We use the symbol $\bbTr{\cdot}$ to denote trace over both classical
and quantum systems, i.e. over both $\mathcal{H}_q$ and
 $\mathcal{H}_c$. We call this bound the Nagaoka--Hayashi
bound. However the Nagaoka--Hayashi bound is not an explicit bound. Our second main result is that this bound, $c_\mathrm{NH}$ can be computed as a semidefinite program.
\subsection{Related bounds}
\label{sec:relatedbounds}
Before proceeding on the proof and computation of the Nagaoka--Hayashi
bound, we digress briefly to mention two related bounds. The first is
the Holevo bound which can be written as~\cite{holevo2011probabilistic}
\begin{align}
\sfTr{\mathsf{V}_\theta}\geq  &\min_{\mathbb{L},\,X}\left\{
                          \bbTr{\mathbb{S}_\theta
                          \mathbb{L}}\,\big|\, \opTr{\mathbb{S}_\theta\mathbb{L}} \text{
       real symmetric, } \opTr{\mathbb{S}_\theta\mathbb{L}}\geq \opTr{\mathbb{S}_\theta{X} X^\intercal},\, \text{ $X_j$
    Hermitian satisfying \eqref{eq:Xunbiased} }
    \right\} \eqqcolon c_\text{H}\;.
\end{align}
As mentioned before, the Holevo bound is a tight bound for collective
measurements in the asymptotic limit. Since the conditions in the
Nagaoka--Hayashi bound $ \mathbb{L}_{jk}=\mathbb{L}_{kj}$ Hermitian
implies $ \opTr{\mathbb{S}_\theta\mathbb{L}}$ real symmetric and
$\mathbb{L}\geq {X} X^\intercal$ implies
$\opTr{\mathbb{S}_\theta\mathbb{L}}\geq \opTr{\mathbb{S}_\theta{X}
  X^\intercal}$, it is clear that the Nagaoka--Hayashi bound is more
restrictive and hence is more informative compared to the Holevo
bound. In other words $c_\text{NH}\geq c_\text{H}$.

The second related bound concerns estimation of physical
observables. In this setting, the operators $X_j$ are given to us as
Hermitian observable operators and the task is to estimate the
expectation values  $\opTr{S_\theta X_j}=x_j $. This
situation is common, for example in state-tomography. Here, in
place of the parameter-MSE matrix~\eqref{eq:parMSE}, we have the
operator-MSE matrix
\begin{align}
  \left[\tilde{\mathsf{U}}_\theta(\Pi,\hat{x})\right]_{jk}= \sum_{m}
  \left(\hat{x}_{jm}-x_j \right)
  \left(\hat{x}_{km}-x_k \right)\opTr{S_\theta
  \Pi_m}\;, \qquad \text{for } j,k=1,2,\dots,n
\end{align}
where we require the classical estimator $\hat{x}$ and POVM
$\Pi$ to satisfy
\begin{align}
  \label{eq:trucond}
  \sum_m {\hat{x}}_{jm} \Pi_m  =X_j\;.
\end{align}
The derivatives of the state $S$ with respect to $\theta$ do not play
any role here. A bound on the trace of $\tilde{\mathsf{U}}_\theta$ is given by
Hayashi's bound~\cite{hayashi1999}
\begin{align}
  \label{eq:gnbU}
\sfTr{\tilde{\mathsf{U}}_\theta}\geq  \min_{\mathbb{L}}\left\{
                          \bbTr{\mathbb{S}_\theta
                          \mathbb{L}}-\sum_jx_j^2\,\big|\, \mathbb{L}_{jk}=\mathbb{L}_{kj} \text{
       Hermitian, } \mathbb{L}\geq {X} X^\intercal
    \right\} \eqqcolon c_\mathrm{NH-U}\;.
\end{align}
As Hayashi's work is only available in Japanese, we summarise its main
results in appendix~\ref{apen:hayashi}. If the given matrices $X$
happen to satisfy the locally unbiasedness
condition~\eqref{eq:Xunbiased} for $\theta$, then $\tilde{\mathsf{U}}_\theta$ also
forms a valid parameter-MSE matrix for those $\theta$. In this case,
because of the additional restriction~\eqref{eq:trucond}, it is clear
that $c_\mathrm{NH-U} \geq\ c_\mathrm{NH}$. Also in this setting,
Watanabe et al.~\cite{watanabe2011uncertainty} derived bounds for estimating two
observables when restricted to certain classes of random and noisy
measurements. In the case when both the observables and state $S$ are
two-dimensional, these bounds are achievable. In fact, when the number
of observables $n=2$, the minimisation over $\mathbb{L}$ can be
performed analytically and $c_\mathrm{NH-U}$ takes the explicit form
\begin{align}
  c_\mathrm{NH-U}=   \opTr{S_\theta X_1 X_1+S_\theta X_2 X_2} +
  \text{TrAbs}\,S_\theta[X_1,X_2]-x_1^2-x_2^2\;.
\end{align}

\subsection{Proof of main results}
\label{sec:construct}
In this section, we shall prove Theorem~\ref{thm:cnh}. To that end, we
need to introduce some definitions. We
rewrite the elements of the MSE matrix as
\begin{align}
\label{mse_def}  \left[\mathsf{V}_\theta\right]_{jk}=
\mathrm{Tr}\bigg[  S_\theta  \underbrace{\sum_{m}\left(\hat{\theta}_{jm}-\theta_j \right)
    \Pi_m \left(\hat{\theta}_{km}-\theta_k \right)}_{[\mathbb{L}_\theta]_{jk}}\bigg]\;,
\end{align}
where the MSE-matrix operator $\mathbb{L}_\theta(\Pi,\hat\theta)$ is an
$n$-by-$n$ matrix with operator elements. We introduce a classical
matrix $\xi$ with elements
$\xi_{jm} \coloneqq \hat\theta_{jm} - \theta_j$ so that
\begin{align}
  \mathbb{L}_\theta
  &=\sum_{m} \begin{pmatrix}
    \xi_{1m}\Pi_m\xi_{1m} & \xi_{1m}\Pi_m\xi_{2m} & \xi_{1m}\Pi_m\xi_{3m} \\
 \xi_{2m}\Pi_m\xi_{1m} & \xi_{2m}\Pi_m\xi_{2m}  &
 \xi_{2m}\Pi_m\xi_{3m} \\
 \xi_{3m}\Pi_m\xi_{1m} & \xi_{3m}\Pi_m\xi_{2m}  & \xi_{3m}\Pi_m\xi_{3m}\end{pmatrix}\\
  &=\sum_{m} \begin{pmatrix}\xi_{1m}\\\xi_{2m}\\\xi_{3m}\end{pmatrix}
  \begin{pmatrix}\xi_{1m}&\xi_{2m}&\xi_{3m}\end{pmatrix} \otimes \Pi_m\;,
\end{align}
where we have set $n=3$ to simplify the presentation. The
generalisation to arbitrary $n$ is straight-forward. With this
notation, it is clear that $\mathbb{L}_\theta$ is an operator on the extended
Hilbert space $\mathcal{H}_c \otimes \mathcal{H}_q$. To anticipate the
proof, it is useful to write $\mathbb{L}_\theta$ in the following form
\begin{align}
\label{eq:mathL}
  \mathbb{L} _\theta
  &=
    \begin{pmatrix}\Xi_{11}&\Xi_{12}&\ldots&\Xi_{1M}\\
      \Xi_{21}&\Xi_{22}&\ldots&\Xi_{2M}\\
      \Xi_{31}&\Xi_{32}&\ldots&\Xi_{3M}
    \end{pmatrix} \begin{pmatrix}\Pi_1&0&\ldots&0\\
      0&\Pi_2&\ldots&0\\
      \vdots&\vdots&\ddots&\vdots\\
      0&0&\ldots&\Pi_M\end{pmatrix}
\begin{pmatrix}\Xi_{11}&\Xi_{21}&\Xi_{31}\\
      \Xi_{12}&\Xi_{22}&\Xi_{32}\\
      \vdots&\vdots&\vdots\\
      \Xi_{1M}&\Xi_{2M}&\Xi_{3M}
    \end{pmatrix} \;,
\end{align}
where $M$ is the number of POVM outcomes and $\Xi_{ij}=\xi_{ij}1$. We can also introduce the
following extension to $S_\theta$,
$ \mathbb{S}_\theta= {1}\otimes S_\theta$ so that the expression for
the MSE matrix can be written as
\begin{align}
  \mathsf{V}_\theta = \mathrm{Tr} \left[\mathbb{S}_\theta \mathbb{L}_\theta \right]\;.
\end{align}
We are now ready to prove Theorem~\ref{thm:cnh}.

\emph{Proof.}
  Suppose the optimal POVM and unbiased estimator have been found and are given by
$\Pi$ and $\hat{\theta}$ which leads to the optimal MSE
\begin{align}
  v^*=\sum_{jm} \xi_{jm}^2 \opTr{S_\theta \,\Pi_m}  =\bbTr{\mathbb{S}_\theta \mathbb{L}^*_\theta} \;.
  \label{voptproof}
\end{align}
We use asterisk to denote the optimal values and optimal operators. From $\Pi$ and $\hat{\theta}$, we can construct the estimator matrices
\begin{align}
  \label{eq:Xstar}
  X_j^* = \sum_m \xi_{jm}\Pi_m\;, \qquad\text{for } j=1,\ldots,n
\end{align}
so that
\begin{align}
  \label{n3cons}
  \begin{pmatrix}X^*_1\\X^*_2\\X^*_3\end{pmatrix}
  \begin{pmatrix}X^*_1&X^*_2&X^*_3\end{pmatrix}
  &=
    \begin{pmatrix}\xi_{11}&\xi_{12}&\ldots&\xi_{1M}\\
      \xi_{21}&\xi_{22}&\ldots&\xi_{2M}\\
      \xi_{31}&\xi_{32}&\ldots&\xi_{3M}
    \end{pmatrix} 
  \begin{pmatrix}\Pi_1\\\Pi_2\\\vdots\\ \Pi_M\end{pmatrix}
  \begin{pmatrix}\Pi_1&\Pi_2&\ldots& \Pi_M
  \end{pmatrix}\begin{pmatrix}\xi_{11}&\xi_{21}&\xi_{31}\\
      \xi_{12}&\xi_{22}&\xi_{32}\\
      \vdots&\vdots&\vdots\\
      \xi_{1M}&\xi_{2M}&\xi_{3M}
    \end{pmatrix}.
\end{align}
Comparing the above with~\eqref{eq:mathL} and using the result
\begin{align}
 \begin{pmatrix}\Pi_1&0&\ldots&0\\
  0&\Pi_2&\ldots&0\\
  \vdots&\vdots&\ddots&\vdots\\
  0&0&\ldots&\Pi_M\end{pmatrix}
               \geq \begin{pmatrix}\Pi_1\\\Pi_2\\\vdots\\ \Pi_M\end{pmatrix}
  \begin{pmatrix}\Pi_1&\Pi_2&\ldots& \Pi_M
  \end{pmatrix}
\end{align}
which holds because $\Pi_j$ are positive operators that sums up to $1$ (see
Proposition II.9.1 of Holevo~\cite{holevo2011probabilistic}), we arrive at
$\mathbb{L}_\theta^* \geq X^*X^{*\intercal}$. With this, we can  bound $v^*$ as
\begin{align}
  v^*&= \bbTr{\mathbb{S}_\theta
                          \mathbb{L}_\theta^*}\\
  &\geq \min_{\mathbb{L}}\left\{
                          \bbTr{\mathbb{S}_\theta
                          \mathbb{L}}\,\big|\, \mathbb{L}_{jk}=\mathbb{L}_{kj} \text{
       Hermitian, } \mathbb{L}\geq {X^*} {X^*}^\intercal
            \right\}\\
  & \geq \min_{\mathbb{L},\,X}\left\{
                          \bbTr{\mathbb{S}_\theta
                          \mathbb{L}}\,\big|\, \mathbb{L}_{jk}=\mathbb{L}_{kj} \text{
       Hermitian, } \mathbb{L}\geq {X} X^\intercal,\, \text{ $X_j$
    Hermitian satisfying \eqref{eq:Xunbiased}}
    \right\} \\&= c_\text{NH}\;. \qed
\end{align}
In the two parameter case, we show in appendix~\ref{app:n=2} that
$c_\text{NH}$ reduces to the original Nagaoka bound $c_{\text{N}}$
in~\eqref{eq_hol2N}. More generally we are interested in minimising
the weighted sum of the covariances which can be formalised with a
positive weight matrix $\mathsf{W} \geq 0$ and minimising
$\sfTr{\mathsf{W}\,\mathsf{V}_\theta}$. This problem can be handled by
a suitable reparametrisation of the model which is presented in
appendix~\ref{apen:weight}.

The Nagaoka--Hayashi bound is not an explicit bound as it still
requires a minimisation over $\mathbb{L}$ and $X$. Our next result
concerns with the computation of this minimisation. Since
$\mathbb{L} - XX^\intercal$ is the Schur's complement of the identity
operator in $ \begin{pmatrix}
  \mathbb{L}&X\\
  X^\intercal&{1}
    \end{pmatrix}
$, the condition  $\mathbb{L} \geq XX^\intercal$ is equivalent to  $  \begin{pmatrix}
    \mathbb{L}&X\\
    X^\intercal&{1}
    \end{pmatrix}\geq 0
$. With this, $c_\mathrm{NH}$ can be written as the semidefinite program
\begin{equation}
  \begin{aligned}
    \label{eq:nh-sdp}
    c_\text{NH} =    \min_{\mathbb{L},\,X}&\,
    \bbTr{\mathbb{S}_\theta
      \mathbb{L}}\;, \\\text{ subject to }&
    \begin{pmatrix}
      \mathbb{L}&X\\
      X^\intercal&{1}
    \end{pmatrix}\geq 0
  \end{aligned}
\end{equation}
where $\mathbb{L}_{jk}=\mathbb{L}_{kj}$ Hermitian and $X_j$ Hermitian
satisfying the conditions~\eqref{eq:Xunbiased} for local
unbiasedness. The conversion to a standard semidefinite program is
performed in appendix~\ref{app:sdp}. We also show in the same appendix
that the worst case computational complexity for solving the SDP to an
accuracy $\epsilon$ is $O\left((nd)^{3/2}\log(1/\epsilon)\right)$.

The computation of the Holevo bound $c_\mathrm{H}$ was shown to be a
semidefinite program by Albarelli et al.~\cite{Albarelli2019}. The difference between
the Holevo bound and the Nagaoka--Hayashi bound is that in the
former, the optimisation is performed directly on the covariance
matrix $\mathsf{V}=\opTr{\mathbb{S}_\theta \mathbb{L}}$ while in the
latter the optimisation is performed on the operators $\mathbb{L}$.
We note that both programs can also be applied to compute
the bound on the operator-MSE $c_\mathrm{NH-U}$~\eqref{eq:gnbU} with
little modification---the only changes needed are to replace the
minimisation variables $X$ with the given observables and ignore the
conditions~\eqref{eq:Xunbiased}.

\section{Examples}
\label{sec:examples}
  In the following, we demonstrate our results by computing the Holevo
and Nagaoka--Hayashi bounds for two illustrative examples---the
estimation of orthogonal qubit rotations on the Bloch sphere in a
phase damping channel and the simultaneous estimation of phase and
loss in an interferometer. In the former we find that the Holevo bound
is always smaller than the Nagaoka--Hayashi bound, and in the latter
we find that the two bounds are always equal. The minimisation problem was
solved with the Yalmip toolbox~\cite{lofberg2004yalmip} for Matlab using
the Mosek solver~\cite{mosek}.

Even though the semidefinite program only returns numerical values for
$X$ and $\mathbb{L}$, in some of these examples, the analytical forms
for them can be inferred from the numerical solutions. Furthermore,
every semidefinite program~\eqref{eq:nh-sdp} has a dual program that
involves performing a maximisation over the Lagrange multipliers
associated with the primal program~\cite{boyd2004}. That the inferred
solutions are indeed optimal can then be verified by checking that the
values for the primal and dual programs coincide. For both of the
examples considered we present the dual solutions in
appendix~\ref{apen:dual}.

\subsection{Example 1: Estimation of qubit rotations with a two-qubit probe}
  \label{sec:examples1}
Our first example concerns estimating the rotation experienced by
qubit probes subject to the phase damping channel. This channel has
particular relevance for modelling decoherence in trapped ions
\cite{huelga1997improvement,myatt2000decoherence,ma2011quantum}. We consider the
maximally entangled two-qubit state
$\left(\ket{01}+\ket{10} \right)/\sqrt{2}$ as a probe. The first qubit acts as a
signal-probe which passes through a channel imparting three small rotations:
$\theta_x$, $\theta_y$ and $\theta_z$ about the $x$, $y$ and $z$ axis
of the Bloch sphere. The rotated probe is then subject to the phase
damping channel $\mathcal{E}$ with a known damping strength $\epsilon$
\begin{equation}
\mathcal{E}[S] =  \left(1-\frac{\epsilon}{2} \right)S+
\frac{\epsilon}{2}\left(\sigma_z\otimes 1 \right)S \left(\sigma_z\otimes 1 \right)\;.
\end{equation}
The second idler-qubit is stored in a perfect quantum memory and
remains unaffected by the rotation or phase damping. The resulting
two-qubit state then has an approximate matrix representation in the computational
basis as
\begin{align}
S_\theta=\frac{1}{4}  \begin{pmatrix}  0&-\I \theta_x -\theta_y&(1-\epsilon)(-\I
  \theta_x-\theta_y)&0\\
  \I \theta_x-\theta_y&2&2(1-\epsilon)(1-\I \theta_z)&(1-\epsilon)(\I
  \theta_x+\theta_y)\\
  (1-\epsilon)(\I \theta_x-\theta_y)&2(1-\epsilon)(1+\I
  \theta_z)&2&\I \theta_x +\theta_y\\
  0&(1-\epsilon)(-\I \theta_x +\theta_y)&-\I \theta_x +\theta_y&0
  \end{pmatrix}\;,
\end{align}
which is valid to the first order in $\theta$. The partial derivatives
of $S_\theta$ with respect to $\theta$ evaluated at $\theta=0$ are
\begin{equation}
  \begin{gathered}
  \frac{\partial S_\theta}{\partial \theta_x}=\frac{1}{4}
  \begin{pmatrix}0&-\I&-\I(1-\epsilon)&0\\
    \I&0&0&\I(1-\epsilon)\\
    \I(1-\epsilon)&0&0&\I\\
    0&-\I(1-\epsilon)&-\I&0
  \end{pmatrix}\;,\\
  \frac{\partial S_\theta}{\partial \theta_y}=\frac{1}{4}
  \begin{pmatrix}0&-1&-(1-\epsilon)&0\\
    -1&0&0&(1-\epsilon)\\
    -(1-\epsilon)&0&0&1\\
    0&(1-\epsilon)&1&0
  \end{pmatrix}\qquad\text{and}\qquad
  \frac{\partial S_\theta}{\partial \theta_z}=\frac{1}{2}
  \begin{pmatrix}0&0&0&0\\
    0&0&-\I(1-\epsilon)&0\\
    0&\I(1-\epsilon)&0&0\\
    0&0&0&0
  \end{pmatrix}\;.
\end{gathered}
\end{equation}

\subsubsection{Single parameter estimation }
Let's start with the simple case when $\theta_y=\theta_z=0$ and we are
only estimating the single parameter $\theta_x$. In a single parameter
estimation problem, the Holevo bound coincides with the
Nagaoka--Hayashi bound and can always be saturated by a separable
measurement. In this case, the two bounds can be achieved by the estimator operator
\begin{equation}
  X_{x}=
  \begin{pmatrix}0&-\I&0&0\\
    \I&0&0&0\\
    0&0&0&\I\\
    0&0&-\I&0
  \end{pmatrix}\;
\end{equation}
which gives $c_{\text{H},1}=c_{\text{NH},1}=1$, independent of $\epsilon$.
The optimal measurement that saturates this bound is a projective
measurement on the four orthogonal eigenvectors of
$X_{x}$
\begin{equation}
  \begin{rcases}\Pi_1\\ \Pi_2
  \end{rcases}=\frac{1}{2}  \begin{pmatrix}1&\mp\I&0&0\\
    \pm\I&1&0&0\\
    0&0&0&0\\
    0&0&0&0
  \end{pmatrix}\;,\qquad
  \begin{rcases}\Pi_3\\ \Pi_4
  \end{rcases}=\frac{1}{2}  \begin{pmatrix}0&0&0&0\\
    0&0&0&0\\
    0&0&1&\mp\I\\
    0&0&\pm\I&1
  \end{pmatrix}\;.
\end{equation}
This together with the estimation coefficients $\xi=(1,-1,-1,1)$ gives
an estimation variance of $v_x=1$. The phase damping channel has no
effect on the estimation precision.

\subsubsection{Two parameter estimation }
Next, for estimating the two parameters $\theta_x$ and $\theta_y$ when
$\theta_z=0$, the Holevo and Nagaoka bounds no longer coincide. The
optimal matrices that achieve the minimum in the Holevo bound are
found to be
\begin{equation}
  \label{hol_Xp2}
  X_{x}=
  \begin{pmatrix}0&-\I&0&0\\
    \I&0&0&0\\
    0&0&0&\I\\
    0&0&-\I&0
  \end{pmatrix}\;,\qquad
    X_{y}=
  \begin{pmatrix}0&-1&0&0\\
    -1&0&0&0\\
    0&0&0&1\\
    0&0&1&0
  \end{pmatrix}\;
\end{equation}
which gives $c_{\text{H},2}=2$. This means that there exists a
sequence of collective measurements that can saturate a variance of
$v_x=v_y=1$ in the asymptotic limit.

Unlike the single parameter case, the optimal $X_{x}$ and
$X_{y}$ operators for the Nagaoka bound are different from those
which optimise the Holevo bound. For the Nagaoka bound the optimal
matrices are
\begin{equation}
  \label{nag_Xp2}
  X_{x}=\frac{1}{2-\epsilon}
  \begin{pmatrix}0&-\I&-\I&0\\
    \I&0&0&\I\\
    \I&0&0&\I\\
    0&-\I&-\I&0
  \end{pmatrix}\;,\qquad
    X_{y}=\frac{1}{2-\epsilon}
  \begin{pmatrix}0&-1&-1&0\\
    -1&0&0&1\\
    -1&0&0&1\\
    0&1&1&0
  \end{pmatrix}\;
\end{equation}
which gives $c_{\text{NH},2}=4/(2-\epsilon)$. Since there is a gap
between the Holevo and Nagaoka bounds, a separable measurement
cannot saturate the Holevo bound---a collective measurement is required. We show in
appendix~\ref{app:phasedampapen} that the Nagaoka bound is saturated
by a family of five-outcome POVMs which gives
$v_x=v_y = 2/(2-\epsilon)$. This means that when restricted to
separable measurements, this is the smallest pair of variances
possible.

\begin{figure}[t!]
\centering
\includegraphics[width=0.7\columnwidth]{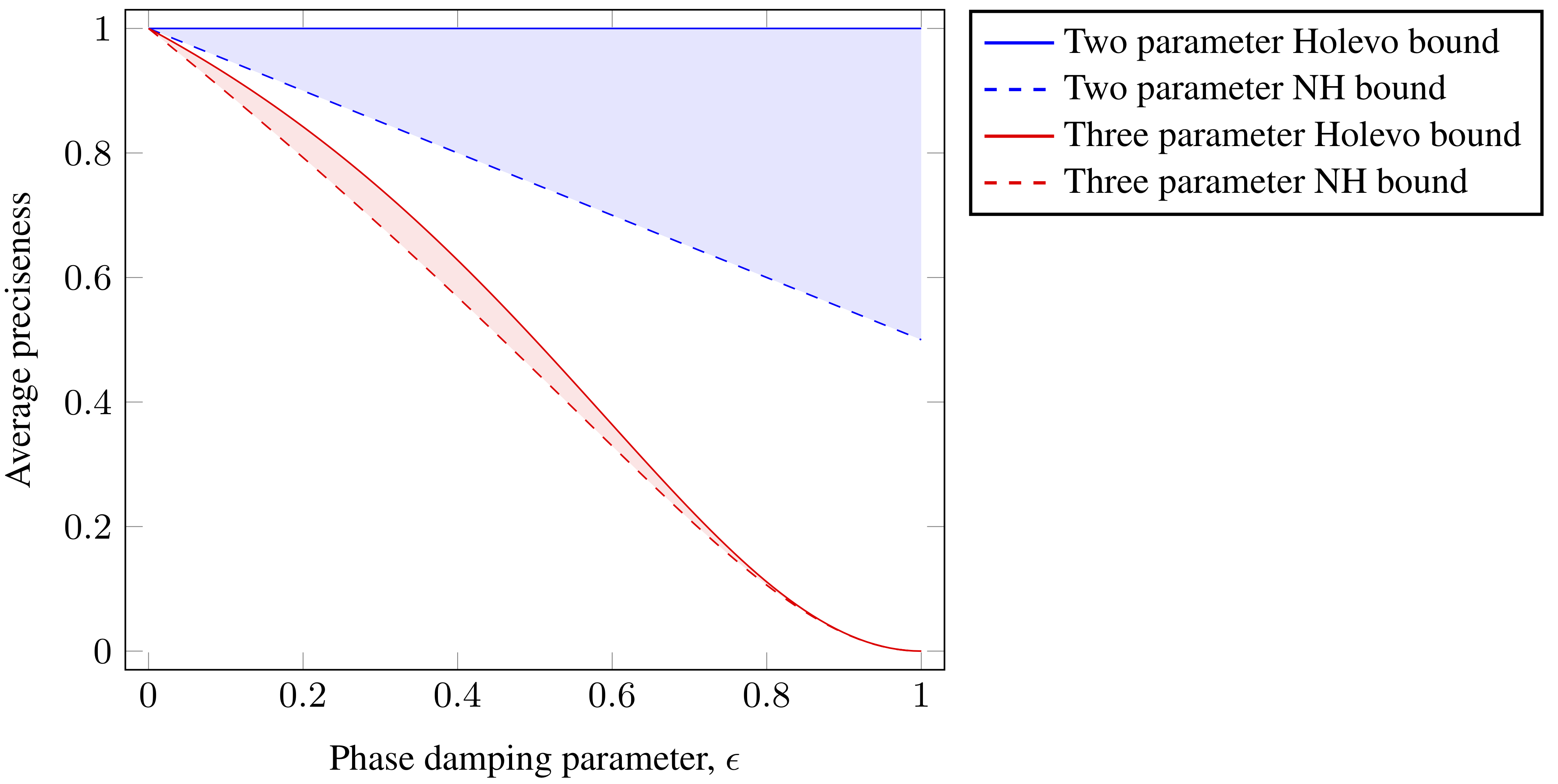}
\caption{Holevo bounds (solid lines) and Nagaoka--Hayashi bounds
  (dashed lines) in terms of average preciseness for estimating two (blue)
  and three (red) orthogonal rotation parameters simultaneously using
  a maximally entangled two-qubit probe under the action of the phase
  damping channel. The Nagaoka--Hayashi bounds can be achieved by a
  separable measurement on a single probe, while the Holevo bound
  require a collective measurement on possibly infinite number of
  copies. The shaded area shows the gap between the two bounds. For
  estimating a single parameter, the Holevo and Nagaoka--Hayashi
  bounds coincide and are equal to the two parameter Holevo bound.}
\label{PDcomp}
\end{figure}
\subsubsection{Three parameter estimation }
Finally for estimating all three angles $\theta_x$, $\theta_y$ and
$\theta_z$ simultaneously we find the Holevo and Nagaoka--Hayashi bounds are
\begin{equation}
c_{\text{H},3}=2+\frac{1}{(1-\epsilon)^2}\qquad\text{ and }\qquad c_{\text{NH},3}=\frac{4}{2-\epsilon}+\frac{1}{(1-\epsilon)^2}\;.
\end{equation} 
Just like the two parameter case, the gap between the two bounds
implies that a collective measurement is required to saturate the
Holevo bound. These bounds are achieved by the same estimator
operators~\eqref{hol_Xp2} for the Holevo bound and~\eqref{nag_Xp2} for
the Nagaoka--Hayashi bound with the additional
\begin{equation}
    X_{z}=\frac{1}{1-\epsilon}
  \begin{pmatrix}0&0&0&0\\
    0&0&-\I&0\\
    0&\I&0&0\\
    0&0&0&0
  \end{pmatrix}\;.
\end{equation}
We write down an explicit POVM that can approach $c_{\text{NH},3}$ with
$v_x=v_y=2/(2-\epsilon)$ and $v_z\rightarrow 1/(1-\epsilon)^2$ in
appendix~\ref{app:phasedampapen} showing that this bound is tight.

In order to quantify the estimation accuracy, we define the
\emph{average preciseness} for simultaneous estimation of $n$
parameters with $n/(v_1+\dots+v_n)$ as a figure of merit on how good
the estimators perform. By construction, a large average preciseness
implies that all $n$ parameters can be determined accurately. We plot
this quantity in Fig.~1 for all three estimation cases. We
also note that in the two and three parameter examples, it is easy to
check that the SLD Fisher information matrix is diagonal. Furthermore
the model is asymptotically classical and the Holevo bound coincides
with the SLD bound~\cite{suzuki2019information,ragy2016compatibility}.

\subsubsection{The Nagaoka--Hayashi bound for multiple copies of the
  probe state}
We now demonstrate the usefulness of the Nagaoka--Hayashi bound and the associated SDP by
examining the precision limits when we perform collective measurements
on finite copies of the probe state. We denote the Nagaoka--Hayashi
bound for $N$ copies of the same probe as
$c_{\text{NH}}(\rho^{\otimes N})$. For a large number of copies of the
probe state we expect the Nagaoka--Hayashi bound to tend to the Holevo
bound,
$\lim\limits_{N\to\infty}Nc_{\text{NH}}(\rho^{\otimes N})= c_{\text{H}}$. For
any finite $N$, we know that
$Nc_{\text{NH}}(\rho^{\otimes N})\geq c_{\text{H}}$ which follows from
$Nc_{\text{H}}(\rho^{\otimes N})=
c_{\text{H}}$. Fig.~2 shows how the gap between
the two bounds shrinks for an increasing number of copies of the probe
state. We consider up to three copies of the probe state. Without the
Nagaoka--Hayashi bound a brute force search for the optimal
measurement strategy for three copies would require optimising an $M$
outcome POVM, where each outcome is a $64$-by-$64$ matrix. Thus the
Nagaoka--Hayashi bound and the associated SDP offer an efficient way to investigate the
asymptotic attainability of the Holevo bound. It provides a tool to
address how fast optimal estimators on finite copies converge to the
asymptotic bound.
\begin{figure}[t!]
\centering
\includegraphics[width=0.85\columnwidth]{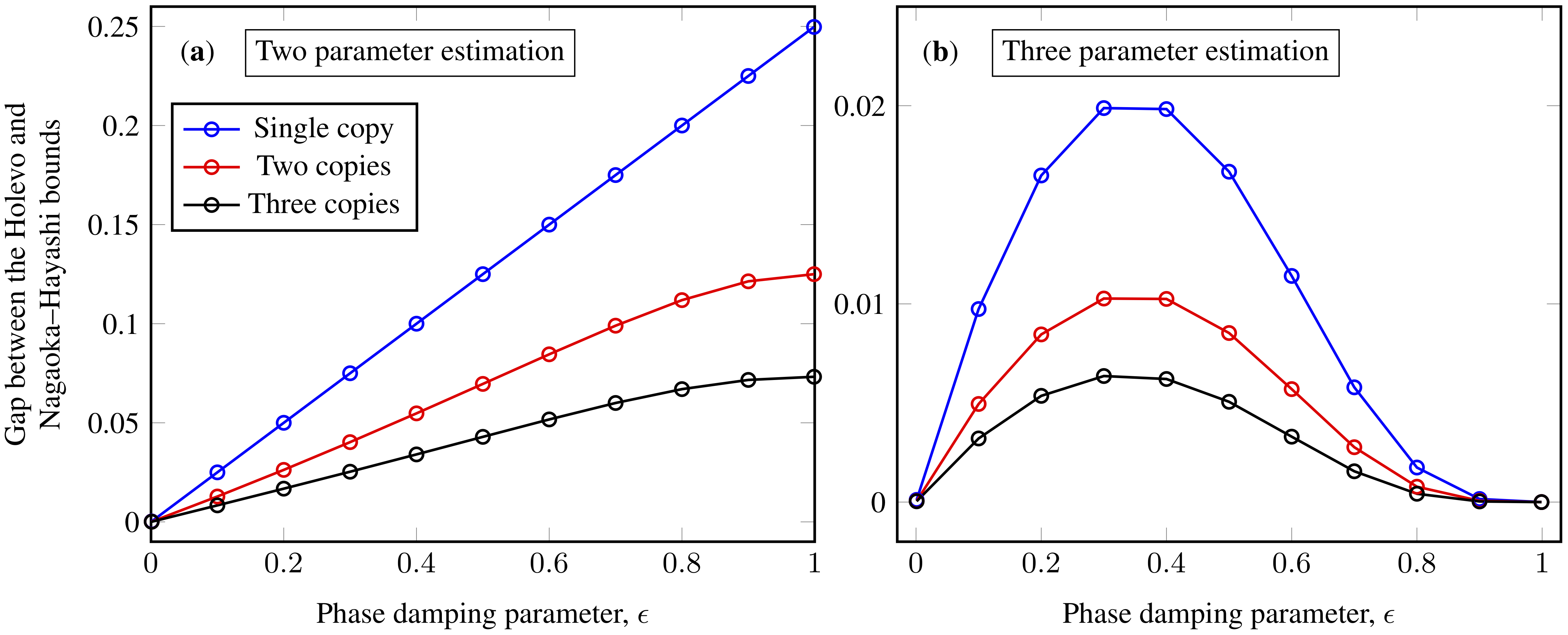}
\caption{Estimating two (a) and three (b) parameters with
 collective measurements on finite copies of the probe state. Both
 figures show how the gap between the Holevo and Nagaoka--Hayashi
 bounds shrinks as the number of copies of the probe state
 increases. The Nagaoka--Hayashi bounds are rescaled by the number of copies
 of the probe state to account for the resources used.}
\label{fig_asympt_attain}
\end{figure}

\subsubsection{Discussion of qubit rotation example}
This example demonstrates several interesting features of finite copy metrology. First we are able to definitively show that there exists a gap between the attainable precision with collective and separable measurements. Without a separable measurement bound such a claim is not possible as any gap between a numerically optimal POVM and the Holevo bound may be a result of a deficiency in the numerical search as opposed to a physically meaningful gap. 
 Secondly as we are able to
 find a POVM which coincides with the Nagaoka--Hayashi bound, we are
 able to say with certainty that this POVM is optimal. 
 Finally we are able to investigate the
 attainability of the Holevo bound. While it is known that the Holevo
 bound is asymptotically attainable, it is not known how many copies
 of the probe state are required to get close to the Holevo bound. As mentioned above to investigate this numerically with a POVM search is computationally very expensive. The SDP presented circumvents this and allows us to investigate the attainability of the Holevo bound in a numerically efficient manner.

\subsection{Example 2: Phase and transmissivity estimation in interferometry}
\label{sec:examples2}
In our next example, we consider the problem of estimation of phase
change $\phi$ and transmissivity $\eta$ in one arm of an interferometer as shown in
Fig.~3. Following Crowley et al.~\cite{Crowley2014}, we consider
initial pure states with a definite photon number $N$ across the two
modes $\ket{\psi_\text{in}}=\sum_{k=0}^N \ket{k,N-k} a_k$, where $\ket{N,M}$ represents a state with $N$ photons in the first mode and $M$ photons in the second mode. One family of states with a fixed photon number is the Holland--Burnett
states which are obtained by interfering two Fock states with an equal
number of photons on a balanced beam splitter. These states lead to a
phase estimation precision better than an interferometer driven by a
coherent light source with the same number of
photons~\cite{Holland1993}. The Holevo bound for the Holland--Burnett
state was computed by Albarelli et al.~\cite{Albarelli2019} for up to $N=14$. In
general, the Holevo bound requires a collective measurement on several
probes to be saturated. But for some values of $N$ and $\eta$, the
Holevo bound can be saturated by a separable measurement, $\Pi^{(\phi)}$ that
optimally measures the phase~\cite{Albarelli2019}.

We compute the Nagaoka bound for these states for different values of
$\eta$ with $\phi=0$ for $N$ up to 14 using our SDP. We find that the Nagaoka and Holevo bounds always
coincide (up to numerical noise). This is to be
expected when $\Pi^{(\phi)}$ saturates the Holevo bound, but is not so
obvious when it does not. The fact that there is no gap between the
Holevo and Nagaoka bound implies one of two possibilities: either (i)
the Nagaoka bound is not tight or (ii) separable measurements are
always optimal for simultaneous estimation of $\phi$ and $\eta$, in
other words, collective measurements cannot do better. In the
following, we show that the second statement is true.
\begin{figure}[t!]
\centering
\includegraphics[width=0.7\columnwidth]{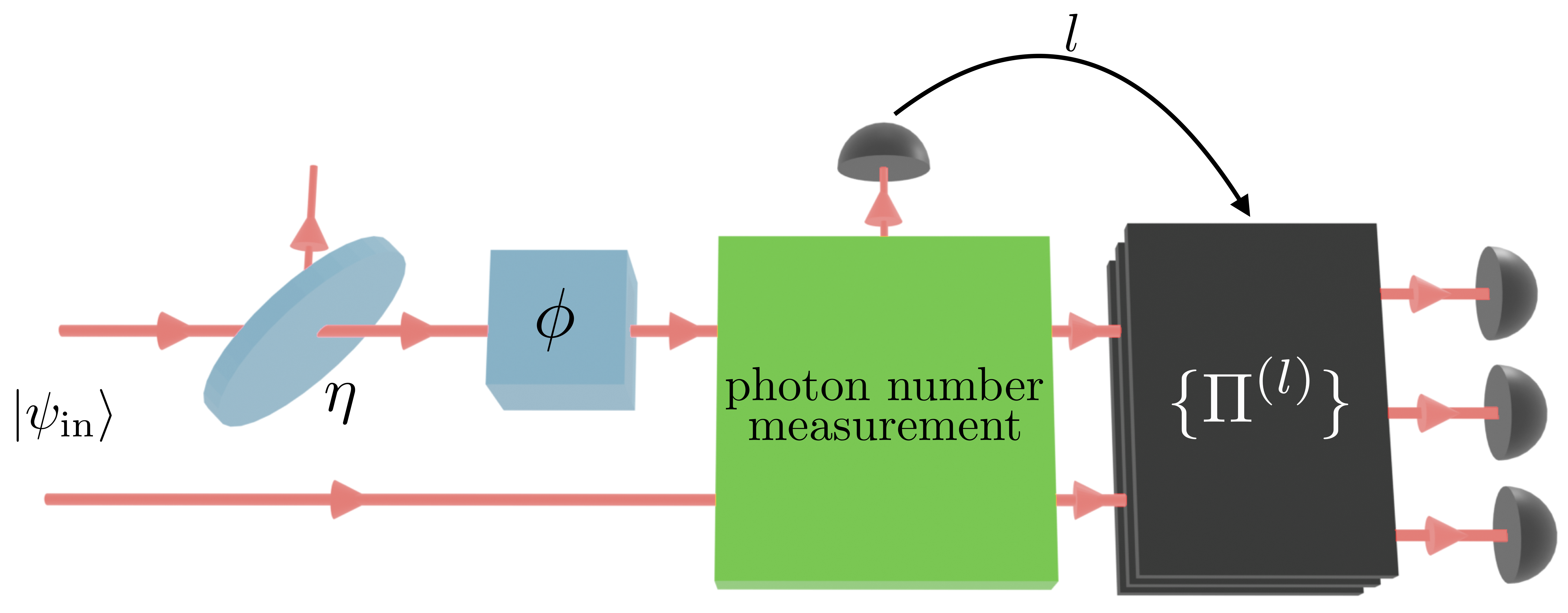}
\caption{\label{fig:oiwl}Schematic for optimal estimation of the phase
  shift $\phi$ and interferometer transmissivity $\eta$ using a two
  mode state $\ket{\psi_\text{in}}$ having definite photon number
  $N$. The measurement can be performed in two stages. The first stage
  (green block) involves performing a projective measurement over the
  photon-number subspace to determine the number of photons lost,
  $l$. The outcome of this measurement is then used to select a
  three-outcome POVM $\{\Pi^{(l)}\}$ for the second stage (black
  box). This measurement strategy saturates not only the Nagaoka
  bound, but also the Holevo bound.}
\end{figure}

\subsubsection{Measurement saturating the Nagaoka bound }
The initial pure state
$\ket{\psi_\text{in}}=\sum_{k=0}^N \ket{k,N-k}a_k$ transforms in
the lossy interferometer channel to the following
state
\begin{align}
  S_{\phi,\eta} = \bigoplus_{l=0}^N \ket{\psi_l} p_l \bra{\psi_l}  \;,
\end{align}
where each term in the direct sum
\begin{align}
  \ket{\psi_l}  =\sum_{k=l}^N \ket{k-l,N-k} a_k e^{\I k \phi}\sqrt{\frac{b_{kl}}{p_l}}
\end{align}
represents a state with $l$ lost photons. The state $S_{\phi,\eta}$ is a mixed state with rank $N+1$. Here
$b_{kl}=\binom{k}{l}\eta^{k-l}(1-\eta)^l$ are the beam-splitter
coefficients and $p_l$ represents the probability of losing $l$
photons. The partial derivatives of $S_{\phi,\eta}$ share the same direct sum
structure
\begin{equation}
  \begin{aligned}
    \frac{\partial S_{\phi,\eta}}{\partial \phi}  &= \bigoplus_{l=0}^N
    \left(\ket{\partial_\phi\psi_l}p_l\bra{\psi_l}+\ket{\psi_l}p_l\bra{\partial_\phi
        \psi_l} \right)\;,\\
    \frac{\partial S_{\phi,\eta}}{\partial \eta}  &= \bigoplus_{l=0}^N\left(
      \ket{\psi_l}\frac{\partial p_l}{\partial
        \eta } \bra{\psi_l}+
      \ket{\partial_\eta\psi_l}p_l\bra{\psi_l}+\ket{\psi_l}p_l\bra{\partial_\eta
        \psi_l} \right)\;,
  \end{aligned}
\end{equation}
with each block having at most rank 2. Thus what we have is a direct sum of
pure state models, and for such a model, we have a separable
measurement with a direct sum structure that can achieve the
Holevo bound~\cite{Matsumoto2002}. Each block can be measured separately
but we cannot minimise $v_\eta+v_\phi$ separately in each block. This
is because how much weight we attach to $\eta$ or $\phi$ in one block
will depend on how much information about them that we can get from
the other blocks. But regardless of the weights, each $l\neq N$ block
requires at most a 3 outcome POVM to saturate the Holevo bound, so the total
number of POVM outcomes needed is at most $3N+1$. The extra 1 comes
from the $l=N$ block where all photons are lost. An analytic POVM that
saturates the Holevo bound for the $N=1$ case is given in
appendix~\ref{app:anal_povm}. The dual solution to the
Nagaoka--Hayashi bound is presented in appendix~\ref{apen:dual_ex2}. 

\subsubsection{Discussion of optical interferometry example}
This problem demonstrates a very different but equally insightful feature of finite copy metrology compared to the qubit rotation problem. The simultaneous estimation of phase and loss has been very well studied in the literature~\cite{dorner2009optimal,demkowicz2009quantum, Crowley2014, Albarelli2019}, however until now the fact that separable measurements are sufficient to reach the ultimate attainable precision had remained unknown.  This insight was only possible with our SDP, which allowed the Nagaoka--Hayashi and Holevo bounds to be compared for large $N$. We plot the numerically calculated Nagaoka--Hayashi and Holevo bounds for different $N$ and $\eta$ in Fig.~4. The fact that collective measurements are not required to reach the Holevo bound in this example may be important from a fundamental viewpoint. 
\begin{figure}[t!]
\centering
\includegraphics[width=0.65\columnwidth]{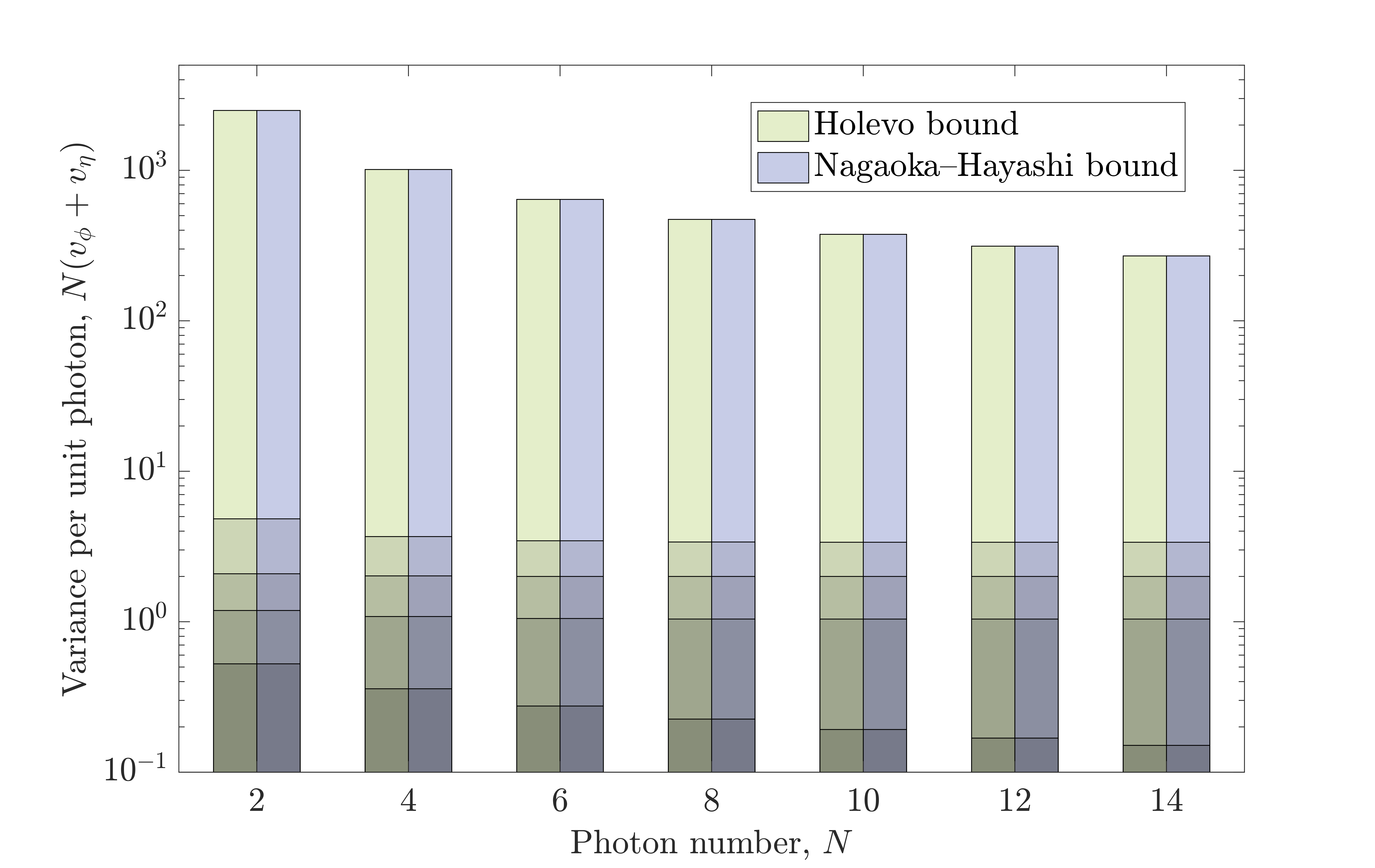}
\caption{\label{fig:oiwlres}The Holevo and Nagaoka--Hayashi bounds per
  unit photon for estimating phase change $\phi$ and transmissivity
  $\eta$ using the Holland--Burnett states. The numerical results
  shows that the two bounds coincide for $N$ up to and including 14
  for any value of $\eta$. We show in the main text that there exists a
  separable measurement that reaches the ultimate attainable precisions
  in this example. In this case, a collective POVM cannot perform
  better than a separable POVM.  Different shades correspond to
  different $\eta$ values, with darker colours corresponding to larger
  $\eta$. Results are shown for $\eta=0.01,0.25,0.5,0.75$ and $0.99$.}
\end{figure}

\section{Discussion}
We have presented the Nagaoka--Hayashi bound for the simultaneous
estimation of multiple parameters when restricted to separable
measurements. This ensures it is always a tighter bound than the
Holevo bound. A gap between the two bounds would imply that the Holevo
bound cannot be achieved with a separable measurement and a collective
measurement is needed to saturate it. Additionally we have shown that
the Nagaoka--Hayashi bound can be formulated as a semidefinite
program, allowing it to be solved efficiently. We have demonstrated
our results with two examples. These examples illustrate how our
results can be used to recognise when a collective measurement is
essential and when it is not. Our results can be applied to
many other problems in multi-parameter quantum metrology and will help
quantify the maximal advantage collective measurements have to offer.
 In some cases, a separable measurement is already optimal, simplifying any experimental realisation.

In the first example, we have assumed that the damping strength
$\epsilon$ is known. However in a practical setting, it would be more
realistic to consider $\epsilon$ as a nuisance parameter, an unknown
parameter that we are not interested in which nevertheless may hinder
our measurement
precision~\cite{suzuki2020nuisance,suzuki2020quantum,tsang2019quantum}. The
quantum Cram\'er--Rao bound in the presence of nuisance parameters can
be computed utilising a low-rank weight
matrix~\cite{suzuki2020nuisance,suzuki2020quantum}.  As we show in
appendix~\ref{apen:weight}, our SDP formalism can be immediately
applied to such cases. An interesting extension to this work would be
to investigate examples which incorporate nuisance parameters.

\section*{Data availability}
The data that supports the findings of this study are available from the corresponding author upon reasonable request.
\section*{Code availability}
The code that supports the findings of this study are available from the corresponding author upon reasonable request.

\section*{Acknowledgement}
This work is supported by the Australian Research Council (ARC) under
the Centre of Excellence for Quantum Computation and Communication
Technology (Grant No. CE170100012). JS is supported by the UEC
Research Support Program, the University of Electro-Communications. We
are grateful to Professor Nagaoka and Professor Hayashi for helpful
discussions.
\section*{Competing Interests}
The authors declare that there are no competing interests.

\section*{Author Contributions}
S.A. and J.S. conceived the project. S.A, J.S. and L.C. developed the
theory, constructed the proof of theorem 1 and worked out the optimal
POVMs in the examples. L.C. and S.A. performed the numerical SDP
simulations. L.C. and S.A. wrote the manuscript and all authors
contributed to discussions regarding the results in this
paper. P.K.L. supervised the project.

\appendix
\section{Summary of Hayashi's results from Ref.~\cite{hayashi1999}}
\label{apen:hayashi}
We summarise Hayashi's result~\cite{hayashi1999}
which was published in the proceedings of a domestic workshop in the Research Institute for Mathematical Sciences (RIMS) at Kyoto University in Japanese for the reader's
convenience. Let ${\cal H}_q$ be a finite $d$-dimensional Hilbert space
and consider a set of observables (Hermitian matrices)
$X=(X_1,X_2,...,X_n)^\intercal$ on it.  We say that a POVM
$\Pi=\{\Pi_m\}$ is a {\it simultaneous measurement} of the given
observables $X$, if
\begin{equation}
X_j=\sum_{m}\hat{x}_{jm}\Pi_m\;,
\end{equation}
holds for all $j$. In general, a projection measurement does not exist
unless the $X_j$ commute with each other, but a POVM $\Pi$ exists.  Given
a state $S$ on ${\cal H}_q$, we define the expectation value of $X_j$ by
\begin{equation}
x_j \coloneqq \opTr{S X_j } = \sum_m\hat{x}_{jm}\opTr{S\Pi_m}\;.
\end{equation}
We define the covariance matrix by 
\begin{align}
\left[\tilde{\mathsf{U}}(\Pi,\hat{x})\right]_{jk}&=\sum_m(\hat{x}_{jm}-
x_j)(\hat{x}_{km}- x_k) \opTr{S \Pi_m} \\
&=\sum_m\hat{x}_{jm}\hat{x}_{km} \opTr{S \Pi_m} -x_jx_k\\
&=\left[\mathsf{U}(\Pi,\hat{x})\right]_{jk} -x_jx_k\;.
\end{align}
We are interested in minimizing the sum of the diagonal elements of $\tilde{\mathsf{U}}(\Pi,\hat{x})$. As the second term is constant this is equivalent to minimising $\sfTr{\mathsf{U}}$. Indeed the second term can be ignored for all practical purposes. We define the precision limit as 
\begin{equation}
C=\inf_\Pi\left\{ \sfTr{\mathsf{U}}-\sum_jx_j^2\,\big|\, \Pi:\text{ simultaneous measurement of }X\right\}\;. 
\end{equation}
Note here that $C$ depends on the given state $S$ and the set of
observables $X$. Hayashi derived the following two bounds for $C$. 
\begin{theorem}[Hayashi]
The following are lower bounds for $C$ and further that $C\ge C_1\ge C_2$ holds.
\begin{align}
C_1&=\inf_{\mathbb{U}}\left\{ \bbTr{\mathbb{U}}-\sum_jx_j^2\,\left|\,\mathbb{U}_{jk}=\mathbb{U}_{kj}\ \mathrm{
       Hermitian},\  
\mathbb{U}\ge\sqrt{\mathbb{S}}XX ^\intercal\sqrt{\mathbb{S}}\right.\right\},\\
C_2&=\inf_\mathsf{U}\left\{ \sfTr{\mathsf{U}}-\sum_jx_j^2\,\left|\,\mathsf{U}\ \mathrm{Hermitian},\ \mathsf{U}\ge\opTr{\sqrt{\mathbb{S}}
XX ^\intercal\sqrt{\mathbb{S}}}\right.\right\}, 
\end{align} 
where $\mathbb{S}=1\otimes S$ and $\mathbb{U}$ are complex matrices on
the extended Hilbert space $\mathcal{H}_c \otimes \mathcal{H}_q$.
\end{theorem}
Hayashi's first bound $C_1$ is considered as the generalisation of the
Nagaoka bound for simultaneous measurement of non-commuting
observables~\cite{nagaoka2005new}. Using the linear programming
approach, Hayashi further derived the following alternative forms for
$C_1$ and $C_2$
\begin{align}
C_1&=\bbTr*{{\rm Sym}_+\left( \sqrt{\mathbb{S}}XX ^\intercal\sqrt{\mathbb{S}}\right)}
+ \inf_{\mathbb{V}}\left\{\bbTr{\mathbb{V}}\,\left|\,\mathbb{V}\ge0,\, 
{\rm Sym}_-\left(\mathbb{V} \right)=- {\rm
     Sym}_-\left(\sqrt{\mathbb{S}}XX ^\intercal\sqrt{\mathbb{S}}
     \right)\right. \right\} -\sum_jx_j^2,\\
C_2&=\bbTr*{{\rm Sym}_+\left( \sqrt{\mathbb{S}}XX ^\intercal\sqrt{\mathbb{S}}\right)}
+ \sfTrAbs*{\opTr*{{\rm Sym}_-\left(\sqrt{\mathbb{S}}XX ^\intercal\sqrt{\mathbb{S}} \right) }}-\sum_jx_j^2\,,
\end{align} 
where ${\rm Sym}_{\pm}(\mathbb{A})=\frac12(\mathbb{A}\pm\mathbb{A}^\intercal) $ is the symmetrised (anti-symmetrized) matrix of $\mathbb{A}$ on $\mathcal{H}_c \otimes \mathcal{H}_q$ with respect to the classical index. 

Finding the fundamental limit $C$ is still an open problem. For two
observables, Nagaoka conjectured that the bound $C_1$ is tight~\cite{nagaoka2005generalization}. In
other words, $C=C_1$.

\section{Nagaoka bound for two parameter estimation}
\label{app:n=2}
The Nagaoka bound for the two parameter estimation case is~\cite{nagaoka2005new}
\begin{align}
  c_{\text{N}} = \min_{X} \big\{ \opTr{S_\theta X_1 X_1+S_\theta X_2 X_2} +
  \text{TrAbs}\,S_\theta[X_1,X_2] \big\}\,
\end{align}
with $X_j$ Hermitian satisfying (4)  in the main text. In this appendix
we show that in the two-parameter case, the Nagaoka--Hayashi
bound, (5) in the main text, coincides with the original Nagaoka bound. When
$n=2$, the Nagaoka--Hayashi bound is
\begin{align}
  \label{eq:b1}
  c_\text{NH} = \min_{\mathbb{L},\,X}\left\{ \bbTr{\mathbb{S}_\theta
  \mathbb{L}}\,\Big|\,\begin{pmatrix}\mathbb{L}_{11}&\mathbb{L}_{12}\\
  \mathbb{L}_{12}&\mathbb{L}_{22}\end{pmatrix} \geq
                   \begin{pmatrix}X_1 X_1&X_1 X_2\\
  X_2 X_1&X_2 X_2\end{pmatrix} \right\}\;,
\end{align}
with $\mathbb{L}_{jk}$ Hermitian and $X_j$ 
Hermitian satisfying (4) in the main text. We can write the condition
  in~\eqref{eq:b1} as
  \begin{align}
  \label{eqb4}
    \begin{pmatrix}\mathbb{L}_{11}&\mathbb{L}_{12}\\
  \mathbb{L}_{12}&\mathbb{L}_{22}\end{pmatrix} -
                   \begin{pmatrix}X_1 X_1&\frac{1}{2}\{X_1,X_2\}\\
                     \frac{1}{2}\{X_2,X_1\}&X_2 X_2\end{pmatrix} &\geq
    \begin{pmatrix}0&\frac{1}{2}[X_1, X_2]\\
      \frac{1}{2}[X_2, X_1]&0\end{pmatrix}\;.
       \end{align}
 Recognising that $[X_1, X_2]/2$ is an antihermitian matrix which we label as $\mathrm{i}H$, we can rewrite the condition as
   \begin{align}
   \label{eqb4true}
    \begin{pmatrix}\mathbb{L}_{11}'&\mathbb{L}_{12}'-\mathrm{i}H\\
  \mathbb{L}_{12}'+\mathrm{i}H&\mathbb{L}_{22}'\end{pmatrix} \geq
   0\;,
       \end{align}
       where $\mathbb{L}'$ denotes the matrix on the left hand side of ~\eqref{eqb4}. In order for this matrix to be positive we require~\cite{lin2015hiroshima}
       \begin{equation}
       \norm{\mathbb{L}_{11}'+\mathbb{L}_{22}'}\geq\norm{2\mathrm{i}H}\;,
       \end{equation}
       for any unitarily invariant norm. This inequality can be saturated by the choice
       \begin{equation}
       \mathbb{L}'= \begin{pmatrix}\abs{H}&0\\
0&\abs{H}\end{pmatrix}\;,
       \end{equation}
       where $\abs{H}=\sqrt{H^2}$. The following corollary ensures~\eqref{eqb4true} is satisfied.
       \begin{corollary}
       Let $A$ be any matrix. Then the matrix $\begin{pmatrix}\abs{A}&A^\dagger\\ A&\abs{A}\end{pmatrix}$ is positive. 
       \end{corollary}
       The matrix $\mathbb{L}'$ can be chosen in this way by optimising over the matrix $\mathbb{L}$ so that
\begin{align}
  \label{eqb4T2}
  \begin{pmatrix}\mathbb{L}_{11}&\mathbb{L}_{12}\\
  \mathbb{L}_{12}&\mathbb{L}_{22}\end{pmatrix} -
                   \begin{pmatrix}X_1 X_1&\frac{1}{2}\{X_1,X_2\}\\
                     \frac{1}{2}\{X_2,X_1\}&X_2 X_2\end{pmatrix} =
    \mathbb{L}'\;,
       \end{align}
hence
\begin{align}
  \label{eqb4T2}
  \min_{\mathbb{L}} \norm{ \begin{pmatrix}\mathbb{L}_{11}&\mathbb{L}_{12}\\
  \mathbb{L}_{12}&\mathbb{L}_{22}\end{pmatrix} -
                   \begin{pmatrix}X_1 X_1&\frac{1}{2}\{X_1,X_2\}\\
                     \frac{1}{2}\{X_2,X_1\}&X_2 X_2\end{pmatrix}} =
   \norm{ \begin{pmatrix}0&\frac{1}{2}[X_1, X_2]\\
      \frac{1}{2}[X_2, X_1]&0\end{pmatrix}}\;.
       \end{align}
       We let this norm be TrAbs, which is equal to the trace for the left hand side of this equation and so the condition becomes
       \begin{equation}
     \min_{\mathbb{L}}  \text{Tr}[\mathbb{L}_{11}+\mathbb{L}_{22}-X_1 X_1-X_2 X_2]=\text{TrAbs}\,2\mathrm{i}H\;.
       \end{equation}
       Rearranging and including $S_\theta$ we arrive at 
       \begin{equation}
       c_\text{NH} =\min_{\mathbb{L},X}\text{Tr}[S_\theta(\mathbb{L}_{11}+\mathbb{L}_{22})]=\min_{X} \text{Tr}[S_\theta( X_1 X_1+ X_2 X_2)]+\text{TrAbs}\,S_\theta[X_1, X_2]=c_\text{N}\;.
       \end{equation}
       
 We now present an alternative method of proving the equivalence between the two bounds for the benefit of the interested reader. We define a complex weight matrix on the extended Hilbert space, whose real part is the identity for simplicity. 
\begin{equation}
\bbW=\sfW\otimes I=\left(\begin{array}{cc}I & -iw I \\ iw I & I\end{array}\right). 
\end{equation}
From positivity of $\sfW$, the parameter $w$ takes values in $[-1,1]$. 
By partial tracing over the parameter space after multiplying $\sqrt{\bbW}$ from both sides, Eq.~\ref{eqb4} implies
\begin{equation}
\sfTr{\sqrt{\bbW}\bbL\sqrt{\bbW}}\ge \sfTr{\sqrt{\bbW} XX^\intercal\sqrt{\bbW}}
\ \Lra\ \bbL_{11}+\bbL_{22}\ge X_1^2+X_2^2+iw[X_1,X_2]. 
\end{equation} 
Since $w$ is arbitrary real in $[-1,1]$, we obtain
\begin{equation}
\bbL_{11}+\bbL_{22} - (X_1^2+X_2^2)\ge\pm i[X_1,X_2]. 
\end{equation}
We then use the following well-known Lemma (see for example Lemma 6.6.1
of Holevo~\cite{holevo2011probabilistic}).
\begin{lemma}
  \label{lem:hol}
For a given Hermitian matrix $\mathsf{Z}$, suppose $\mathsf{Y}$ obeys inequalities $\mathsf{Y}\ge \pm \mathsf{Z}$. 
Then, the minimum of the trace of $\mathsf{Y}$ is given by
\[
\min_{\mathsf{Y}}\left\{ \sfTr{\mathsf{Y}}  \,\big|\, \mathsf{Y} \;\mathrm{Hermitian} ,\,
\mathsf{Y} \ge \pm \mathsf{Z} \right\}=\sfTrAbs{\mathsf{Z}}\;.
\]
\end{lemma}
This gives the bound
\begin{align}
\label{eqb14}
 &\opTr{S_\theta(\bbL_{11}+\bbL_{22})}-\opTr{S_\theta(X_1^2+X_2^2)}\ge \mathrm{TrAbs} \sqrt{S_\theta}[X_1,X_2]\sqrt{S_\theta} \\
\Lra\ & \opTr{S_\theta(\bbL_{11}+\bbL_{22})}\ge 
\opTr{S_\theta(X_1^2+X_2^2)}+ \mathrm{TrAbs} \sqrt{S_\theta}[X_1,X_2]\sqrt{S_\theta}
\end{align}
As before the left hand side of Eq.~\ref{eqb14} can be chosen so that the two sides are equal. Thus, $\min \bbTr{\bbS_\theta\bbL}$ under the constraints is given by 
\begin{equation}
\min_{X} \opTr{S_\theta(X_1^2+X_2^2)}+ \mathrm{Tr}{\left| \sqrt{S_\theta}[X_1,X_2]\sqrt{S_\theta}\right|}.
\end{equation}

Note that this method works for more general weight matrices. 


\section{Generalisation to arbitrary weight matrix}
\label{apen:weight}
We present a generalisation of our main results to
an arbitrary weight matrix $\mathsf{W}\geq 0$. In the case where the
weight matrix $\mathsf{W}>0$ is full rank, it can be set to the
identity after a suitable reparametrisation for the model (see for
example, Sec. V of Fujiwara and Nagaoka~\cite{fujiwara1999estimation}). Since we are only
interested in local bound, this reparametrisation does not
matter. Specifically, we can reparametrise the model as
$\varphi_j=\sum_k \mathsf{H}_{jk} \theta_k$ where
$\mathsf{H}=\sqrt{\mathsf{W}}$ is a real and regular matrix. Estimating the new
parameters $\varphi$ is equivalent to estimating the original
parameters $\theta$ with a weight matrix $\mathsf{W}$.

When $\mathsf{W}$ is not full rank, a bit more care is required in
reparametrising the model because it might be possible that some of
the new parameters $\varphi_j$ are exactly zero or that two of the
$\varphi_j$'s might be identical. This situation is common when
studying parameter estimation in the presence of nuisance
parameters~\cite{suzuki2020nuisance,suzuki2020quantum,tsang2019quantum}. Nonetheless,
it is still easy to incorporate the weight matrix $\mathsf{W}$ into
our original framework. We now wish to minimise
$\sfTr{\mathsf{W}\,\mathsf{V}_\theta}$ instead of
$\sfTr{\mathsf{V}_\theta}$.  Recalling that the MSE matrix can be
written as
$\mathsf{V}_\theta = \opTr{\mathbb{S}_\theta \mathbb{L}_\theta }$,
this is handled by noting the following
\begin{align}
  \mathsf{W}\,\mathsf{V}_\theta
  &=\mathsf{W} \,\opTr*{\mathbb{S}_\theta \mathbb{L}_\theta }\\
  &=\opTr*{ \left({\mathsf{W}}\otimes 1\right)\mathbb{S}_\theta
    \mathbb{L}_\theta } \\
  &=\opTr*{\mathbb{S}'_\theta \mathbb{L}_\theta}\;,
\end{align}
where
$\mathbb{S}'_\theta=(\mathsf{W}\otimes
1)\mathbb{S}_\theta=\mathsf{W} \otimes S_\theta$ is a positive
semidefinite matrix. Thus, by changing from
$\mathbb{S}_\theta$ to $\mathbb{S}'_\theta$, nothing about the problem
changes and it can be solved using the same SDP as in the main text.

\section{Conversion to standard SDP and complexity discussions}
\label{app:sdp}
Here we show that the program
\begin{equation}
  \label{eq:primary_problem1}
  \begin{aligned}
  c_\text{NH} =    \min_{\mathbb{L},\,X}&\,
                          \bbTr{\mathbb{S}_\theta
  \mathbb{L}}\;,\\ \text{ subject to }&
  \begin{pmatrix}
    \mathbb{L}&X\\
    X^\intercal&1
    \end{pmatrix}\geq 0
  \end{aligned}
\end{equation}
with $\mathbb{L}_{jk}=\mathbb{L}_{kj}$ Hermitian and $X_j$ Hermitian
satisfying (4) in the main text can be converted to the standard SDP
program
\begin{equation}
  \label{eq:primary_problem2}
\begin{gathered}
c_\text{NH} = \min_{Y \geq 0 }\, \bbTr{ \bm{F}_0 \bm{Y}}  \\
\text{subject to } \bbTr{ \bm{F}_k \bm{Y}} =c_k\;,\qquad\text{for }k=1,\ldots,m\;,
\end{gathered}
\end{equation}
where $\bm{Y}$ is a positive-semidefinite Hermitian matrix of size $nd+d$ having the form
$\bm{Y}= \begin{pmatrix}
    \mathbb{L}&X\\
    X^\intercal&1
    \end{pmatrix}$, $d$ is the dimension of $\mathcal{H}_q$ and $m$ is
    the total number of constraints on $\bm{Y}$. The objective function
to be minimised is handled with
\begin{align}
  \bm{F}_0 = \begin{pmatrix}\mathbb{S}_\theta&0\\0&0\end{pmatrix}  \;.
\end{align}
There are five groups of constraints on $\bm{Y}$ that have to be implemented
through $\bm{F}_k$ and $c_k$. Denoting $S_j=\frac{\partial S_\theta}{\partial\theta_j}$, the constraints are:
\begin{enumerate}
\item $\opTr{S_\theta X_j} = \theta_j$.
\item $\opTr{S_j X_k} = \delta_{jk}$.
\item $X_j$ Hermitian.
\item $\mathbb{L}_{jk}=\mathbb{L}_{kj}$ Hermitian.
\item The lower $n$-by-$n$ block of $\bm{Y}$ equals the identity operator.
\end{enumerate}
In the following, we set $n=3$ to simplify the notations. The group 1
constraints are achieved with the $n$ matrices and
constants
\begin{equation}
  \begin{aligned}
  \bm{F}^{(1)}_{1} &= \begin{pmatrix} 0& \begin{pmatrix}S_\theta\\0\\0\end{pmatrix}\\
     \begin{pmatrix}S_\theta&0&0\end{pmatrix}&0
   \end{pmatrix},\qquad c^{(1)}_{1}=2 \theta_1\;,\\
    \bm{F}^{(1)}_{2} &= \begin{pmatrix} 0& \begin{pmatrix}0\\S_\theta\\0\end{pmatrix}\\
     \begin{pmatrix}0&S_\theta&0\end{pmatrix}&0
  \end{pmatrix},\qquad c^{(1)}_{2}=2 \theta_2\;,\\
  \bm{F}^{(1)}_{3} &= \begin{pmatrix} 0& \begin{pmatrix}0\\0\\S_\theta\end{pmatrix}\\
     \begin{pmatrix}0&0&S_\theta\end{pmatrix}&0
  \end{pmatrix},\qquad c^{(1)}_{3}=2 \theta_3\;.
\end{aligned}
\end{equation}
The group 2
constraints are achieved with the $n{\times} n$ matrices and
constants
\begin{equation}
\begin{aligned}
  \bm{F}^{(2)}_{1j} &= \begin{pmatrix} 0& \begin{pmatrix}S_j\\0\\0\end{pmatrix}\\
     \begin{pmatrix}S_j&0&0\end{pmatrix}&0
  \end{pmatrix},\qquad c^{(2)}_{1j}=2 \delta_{1j}\;,\\
  \bm{F}^{(2)}_{2j} &= \begin{pmatrix} 0& \begin{pmatrix}0\\S_j\\0\end{pmatrix}\\
     \begin{pmatrix}0&S_j&0\end{pmatrix}&0
  \end{pmatrix},\qquad c^{(2)}_{2j}=2 \delta_{2j}\;,\\
  \bm{F}^{(2)}_{3j} &= \begin{pmatrix} 0& \begin{pmatrix}0\\0\\S_j\end{pmatrix}\\
     \begin{pmatrix}0&0&S_j\end{pmatrix}&0
  \end{pmatrix},\qquad c^{(2)}_{3j}=2 \delta_{3j}\;,
\end{aligned}
\end{equation}
for $j=1,\ldots,n$. To implement the rest of the constraints, we
introduce $d^2$ Hermitian basis-operators $B_j$ for
$\mathcal{L}(\mathcal{H}_q)$ where $\mathcal{L}(\mathcal{H}_q)$ denote the space of Hermitian operators
in $\mathcal{H}_q$, $\opTr{B_j B_k}=\delta_{jk}$ and $B_1$ proportional to the
identity~\cite{hioe1981,kimura2003,bertlmann2008}.  If $S_\theta$ is not full rank, the number of
basis operators can be reduced by $(d-r)^2$ where $r$ is the rank of
$S_\theta$ by restricting $B_j$ to the quotient space
$\mathcal{L}(\mathcal{H}_q)/\mathcal{L}(\text{ker}(S_\theta))$. See for
example the discussions in~\cite[Sec.~2.10]{holevo2011probabilistic}
or~\cite{Albarelli2019}. The group 3 constraints are then implemented
by $n{\times} d^2$ matrices and constants
\begin{equation}
  \begin{aligned}
  \bm{F}^{(3)}_{1j} &= \begin{pmatrix} 0& \begin{pmatrix} \I B_j\\0\\0\end{pmatrix}\\
     \begin{pmatrix}-\I B_j&0&0\end{pmatrix}&0
  \end{pmatrix},\qquad c^{(3)}_{1j}=0\;,\\
  \bm{F}^{(3)}_{2j} &= \begin{pmatrix} 0& \begin{pmatrix}0\\\I B_j\\0\end{pmatrix}\\
     \begin{pmatrix}0&-\I B_j&0\end{pmatrix}&0
  \end{pmatrix},\qquad c^{(3)}_{2j}=0\;,\\
  \bm{F}^{(3)}_{3j} &= \begin{pmatrix} 0& \begin{pmatrix}0\\0\\ \I B_j\end{pmatrix}\\
     \begin{pmatrix}0&0& -\I B_j\end{pmatrix}&0
  \end{pmatrix},\qquad c^{(3)}_{3j}=0\;,
\end{aligned}
\end{equation}
for $j=1,\ldots,d^2$. The group 4 constraints are implemented
with $\dfrac{n^2-n}{2}{\times}d^2$ matrices and constants
\begin{equation}
\begin{aligned}
  \bm{F}^{(4)}_{1,2,j} &= \begin{pmatrix}
    \begin{pmatrix} 0&\I B_j&0\\
      -\I B_j&0&0\\
    0&0&0\end{pmatrix}&0\\
     0&0
  \end{pmatrix},\qquad c^{(4)}_{1,2,j}=0\;,\\
  \bm{F}^{(4)}_{1,3,j} &= \begin{pmatrix}
    \begin{pmatrix} 0&0&\I B_j\\
      0&0&0\\
    -\I B_j&0&0
    \end{pmatrix}&0\\
     0&0
  \end{pmatrix},\qquad c^{(4)}_{1,3,j}=0\;,\\
  \bm{F}^{(4)}_{2,3,j} &= \begin{pmatrix}
    \begin{pmatrix} 0&0&0\\
      0&0&\I B_j\\
    0&-\I B_j&0\end{pmatrix}&0\\
     0&0
  \end{pmatrix},\qquad c^{(4)}_{2,3,j}=0\;.
\end{aligned}
\end{equation}
for $j=1,\ldots,d^2$. Finally, the group 5 constraints are implemented
with $d^2$ matrices and constants
\begin{align}
    \bm{F}^{(5)}_{1} = \begin{pmatrix}
      0&0\\
      0&B_1
    \end{pmatrix},\qquad c^{(5)}_{1}=\sqrt{d},\qquad\text{and}\qquad
    \bm{F}^{(5)}_{j} = \begin{pmatrix}
      0&0\\
      0&B_j
  \end{pmatrix},\qquad c^{(5)}_{j}=0
\end{align}
for $j=2,3,\ldots,d^2$.

The worst-case time complexity for solving the
SDP~(\ref{eq:primary_problem1}) or~(\ref{eq:primary_problem2}) to a
desired accuracy $\epsilon$ is $O(\sqrt{N}\log(1/\epsilon))$ where
$N=(n+1)d$ is the size of the matrix
$\bm{F}_0$~\cite{boyd2004,vandenberghe1996semidefinite}. However in
our simulations, we observed that the time complexity is independent
of $N$. This is consistent with reports in the literature that in
practice, the SDP algorithms perform much better than its worst-case
bound~\cite{vandenberghe1996semidefinite}. Each time step requires
solving a system of linear equations with a computational complexity
of $O(N^3)$. Therefore, the overall worst-case computational complexity is
$O\left(N^{3/2}\log(1/\epsilon)\right)$.

\section{Estimation of qubit rotations under phase damping channel
  with a two-qubit probe---analytic POVM saturating the
  Nagaoka--Hayashi bound}
\label{app:phasedampapen}
We now present an analytic measurement strategy that
saturates the Nagaoka--Hayashi bound for the qubit rotation estimation
problem. We first define the four sub-normalised projectors
\begin{align}
  \begin{drcases}\ket{\phi_1}\\
    \ket{\phi_2}\end{drcases}=\frac{1}{2}
  \begin{pmatrix}1\\\pm a\I\\\pm a\I\\1\end{pmatrix}\qquad \text{and}\qquad
  \begin{drcases}\ket{\phi_3}\\
    \ket{\phi_4}\end{drcases}=\frac{1}{2}
  \begin{pmatrix}1\\\mp b\\\mp b\\-1\end{pmatrix}\;
\end{align}
where $a$ and $b$ are two non-zero real parameters satisfying
$a^{2}+b^{2}\leq 1$. An optimal strategy that saturates the Nagaoka
bound for estimating $\theta_x$ and $\theta_y$ consists of measuring
the five-outcome POVM with $\Pi_j = \ketbra{\phi_j}$ for $j=1,2,3,4$
and $\Pi_{5}=1-(\Pi_1 + \Pi_2 + \Pi_3 +\Pi_4)$.  The probability for
each POVM outcome is
\begin{equation}
\begin{aligned}
  \begin{drcases}p_1\\
    p_2\end{drcases}&=\frac{1}{4}a(2-\epsilon)(a\pm \theta_x)\;,\\
  \begin{drcases}p_3\\
    p_4\end{drcases}&=\frac{1}{4}b(2-\epsilon)(b\pm \theta_y)\;,\\
  p_5&=1- \frac{1}{2}(2-\epsilon)(a^2+b^2)\;.
\end{aligned}
\end{equation}
We can use this to construct unbiased estimators for $\theta_{x}$ and $\theta_{y}$ with
\begin{equation}
  \begin{gathered}
\xi_{x,1} = -\xi_{x,2} =\frac{2}{(2-\epsilon)a} \;,\qquad\xi_{x,3}=\xi_{x,4}=\xi_{x,5}=0\;,\\
\xi_{y,3}=-\xi_{y,4}=\frac{2}{(2-\epsilon)b}\;,\qquad \xi_{y,1}=\xi_{y,2}=\xi_{y,5}=0\;.
\end{gathered}
\end{equation}
In this construction, the fifth outcome $\Pi_5$ does not give any
additional information about $\theta_x$ or $\theta_y$. Nonetheless, it
is still necessary to be included so that the POVM outcomes sum up to
1. For a finite sample, to have a better estimate of $\theta_x$ and
$\theta_y$, it is thus beneficial to have both $a$ and $b$ large so
the outcomes $\Pi_1$ to $\Pi_4$ occur more often. However, in the
asymptotic limit, the variances in our estimate of $\theta_{x}$ and
$\theta_{y}$ are
\begin{equation}
  \begin{aligned}
    v_{x}&=\xi_{x,1}^2 \,p_1 + \xi_{x,2}^2\, p_2
    =\frac{4(p_1+p_2)}{(2-\epsilon)^2 a^2}=\frac{2}{2-\epsilon}\;,\\
    v_{y}&=\xi_{y,3}^2 \,p_3 + \xi_{y,4}^2\,
    p_4=\frac{4(p_3+p_4)}{(2-\epsilon)^2 b^2}=\frac{2}{2-\epsilon}
  \end{aligned}
\end{equation}
which do not depend on $a$ or $b$. The sum $v_x+v_y=4/(2-\epsilon)$
saturates the Nagaoka bound as claimed.

For estimating all three parameters $\theta_x$, $\theta_y$ and
$\theta_z$, one measurement strategy is to
use the same POVM outcomes for estimating
$\theta_x$ and $\theta_y$ but splitting $\Pi_5$ to get some
information on $\theta_z$. Ideally, we would like to use these four
projectors we get when setting $a=b=0$,
\begin{align}
  \label{nag_2pp}
  \Pi_1=\Pi_2&=\frac{1}{4}\begin{pmatrix}1&0&0&1\\
    0&0&0&0\\
    0&0&0&0\\
    1&0&0&1
  \end{pmatrix},\qquad
  \Pi_3=\Pi_4=\frac{1}{4}\begin{pmatrix}1&0&0&-1\\
    0&0&0&0\\
    0&0&0&0\\
    -1&0&0&1
  \end{pmatrix},\qquad
\Pi_5= \frac{1}{2}\begin{pmatrix}0&0&0&0\\
    0&1&\I&0\\
    0&-\I&1&0\\
    0&0&0&0
  \end{pmatrix}+
    \frac{1}{2}\begin{pmatrix}0&0&0&0\\
    0&1&-\I&0\\
    0&\I&1&0\\
    0&0&0&0
  \end{pmatrix}
\end{align}
to obtain the most information on $\theta_z$ without affecting the estimate
of $\theta_x$ and $\theta_y$. But the problem is that at this
singular point, the first
four outcomes $\Pi_1$, $\Pi_2$, $\Pi_3$ and $\Pi_4$ do not give any
information on $\theta_x$ and $\theta_y$. To fix this, we need both $a$
and $b$ to be close to but not exactly zero. Writing
$\delta=(a^2+b^2)/2$, we can split $\Pi_5$ as
\begin{align}
  \Pi_{5}&=\begin{pmatrix}0&0&0&0\\
    0&1-\delta&-\delta&0\\
    0&-\delta&1-\delta&0\\
    0&0&0&0
  \end{pmatrix}\\
  &=\underbrace{\delta\begin{pmatrix}0&0&0&0\\
    0&1&-1&0\\
    0&-1&1&0\\
    0&0&0&0
  \end{pmatrix}}_{\Pi_5^{(3)}}+
\underbrace{          \frac{1-2\delta}{2}\begin{pmatrix}
             0&0&0&0\\
             0&1&-\I&0\\
             0&\I&1&0\\
             0&0&0&0
  \end{pmatrix}}_{\Pi_6^{(3)}}+
\underbrace{           \frac{1-2\delta}{2}\begin{pmatrix}0&0&0&0\\
    0&1&\I&0\\
    0&-\I&1&0\\
    0&0&0&0
  \end{pmatrix}}_{\Pi_7^{(3)}}
\end{align}
which has outcome probabilities
\begin{equation}
\begin{aligned}
  p_5&=\delta \,\epsilon\;,\\
  \begin{drcases}p_6\\
    p_7\end{drcases}&=\frac{1}{2}(1-2\delta)\left(1\pm (1-\epsilon
    )\theta_z \right)\;.
\end{aligned}
\end{equation}
This together with
\begin{equation}
  \xi_{z,1}=\xi_{z,2}=\xi_{z,3}=\xi_{z,4}=\xi_{z,5}=0\;,\qquad
  \text{and}\qquad \xi_{z,6}=-\xi_{z,7}=\frac{1}{(1-\epsilon)(1-2\delta)}\;,
\end{equation}
give a variance for estimating $\theta_z$ as
$  v_z=\dfrac{1}{(1-\epsilon)^2(1-2\delta)} $
which approaches $v_z=\dfrac{1}{(1-\epsilon)^2}$ as $\delta$ tends to zero. 

\section{Phase and transmissivity estimation in
  interferometry---analytic POVM saturating the Holevo Cram\'er--Rao
  bound for 1 photon state}
\label{app:anal_povm}
Consider the 1 photon state
$\ket{\psi_\text{in}}=\ket{01}a_0+\ket{10}a_1$ where $a_0$ and $a_1$
are positive coefficients. This state transforms
through the lossy interferometer with transmissivity $\eta$ and a
phase shift $\phi$ to the state with matrix representation
\begin{align}
S_\theta=  \begin{pmatrix}
  (1-\eta)a_1^2&0&0\\
  0&a_0^2&\sqrt{\eta} a_0 a_1 e^{-\I\phi}\\
   0&\sqrt{\eta} a_0 a_1 e^{\I\phi}& \eta a_1^2\\
  \end{pmatrix}\,
\end{align}
whose derivatives evaluated at $\phi=0$ are
\begin{align}
\frac{\partial S_\theta}{\partial \eta}=  \begin{pmatrix}
  -a_1^2&0 &0\\
  0&0&\frac{a_0 a_1}{2\sqrt{\eta}}\\
  0&\frac{a_0 a_1}{2\sqrt{\eta}}&a_1^2
  \end{pmatrix}\qquad\text{and}\qquad
\frac{\partial S_\theta}{\partial \phi}=  \begin{pmatrix}
  0&0 &0\\
  0&0&-\I \sqrt{\eta}a_0 a_1\\
  0&\I \sqrt{\eta}a_0 a_1&0
\end{pmatrix}\;,
\end{align}
where the matrix basis is $\left\{\ket{00},\ket{01},\ket{10}
\right\}$. The Holevo bound for this model was computed
by Albarelli et al.~\cite{Albarelli2019} to be
\begin{align}
  \label{n1HCR}
  c_\text{H}= \begin{dcases}
        \frac{1+3\eta-4\eta^3}{4\eta a_1^2} \qquad&\text{for } a_1 < \frac{1}{\sqrt{2}} \text{ and }
        \eta < \frac{a_0^2-a_1^2}{2a_0^2}\;,\\
            \frac{\left(a_0^2 +\eta
        a_1^2\right)\left(1+4\eta(1-\eta)a_0^2\right)}{4 \eta a_0^2
      a_1^2 }\qquad &\text{otherwise. }
  \end{dcases}
\end{align}
In the following, we show that this bound can be saturated by a
separable measurement. There exist a family of measurements that 
can saturate the Holevo bound. One of them is the four-outcome
POVM
\begin{equation}
  \begin{aligned}
  \label{eq:4pi}
  \Pi_1  &=
  \begin{pmatrix}
    1&0&0\\
    0&0&0\\
    0&0&0
  \end{pmatrix}\;,\qquad
  \Pi_2  =
  \begin{pmatrix}
    0&0&0\\
    0&0&0\\
    0&0&1-\dfrac{a_0^2}{(1-\eta)(1+2\eta)a_0^2 - \eta a_1^2}
  \end{pmatrix}\;,\\
  \begin{drcases}\Pi_{3}\\ \Pi_4\end{drcases}&=\frac{1}{2}
  \begin{pmatrix}
    0&0&0\\
    0&1&\mp \dfrac{\I a_0}{\sqrt{(1-\eta)(1+2\eta)a_0^2-\eta a_1^2}}\\
    0&\pm  \dfrac{\I a_0}{\sqrt{(1-\eta)(1+2\eta)a_0^2-\eta a_1^2}}& \dfrac{a_0^2}{(1-\eta)(1+2\eta)a_0^2-\eta a_1^2}
  \end{pmatrix}  
\end{aligned}
\end{equation}
together with the estimation coefficients
\begin{equation}
\begin{aligned}
\xi_{\eta,1}&=-\frac{1+2\eta}{2a_1^2}\;,\qquad
  \xi_{\eta,2}=\frac{(1-\eta)(1+2\eta)}{2\eta a_1^2}\;,\qquad\xi_{\eta,3}=\xi_{\eta,4}=\frac{1}{2a_0^2}\;,
              \\
\xi_{\phi,1}&=\xi_{\phi,2}=0\;,\qquad\text{and}\qquad\xi_{\phi,3}=-\xi_{\phi4}= \frac{\sqrt{(1-\eta)(1+2\eta)a_0^2-\eta a_1^2}}{2\sqrt{\eta}a_0^2 a_1}\;.
\end{aligned}
\end{equation}
One can verify that when $\eta < (a_0^2-a_1^2)/2$, these outcomes are
non-negative operators that satisfy $\Pi_1+\Pi_2+\Pi_3+\Pi_4=1$. The
estimator matrices
$ X_\eta = \xi_{\eta,1}\Pi_1 + \xi_{\eta,2}\Pi_2 + \xi_{\eta,3}\Pi_3 +
\xi_{\eta,4}\Pi_4\;\text{ and }\; X_\phi = \xi_{\phi, 3}\Pi_3 + \xi_{
  \phi, 4}\Pi_4 $ satisfy the unbiased
conditions, (4) in the main text. The probability for each outcome to
occur is
\begin{equation}
  \begin{aligned}
    p_1&=(1-\eta)a_1^2\;,\\
    p_2&=\eta a_1^2- \frac{\eta a_0^2 a_1^2}{(1-\eta)(1+2\eta)-\eta a_1^2}\;,\\
    p_3&=p_4=\frac{a_0^2}{2}\left(1+ \frac{\eta
        a_1}{(1-\eta)(1+2\eta)a_0^2-\eta a_1^2}\right)\;.
  \end{aligned}
\end{equation}
The variances of these two estimators are
\begin{equation}
\begin{aligned}
  v_\eta &= \xi_{\eta,1}^2 \,p_1 + \xi_{\eta,2}^2 \,p_2+ \xi_{\eta,3}^2 \,p_3 + \xi_{\eta,4}^2 \,p_4
  = \frac{1+\eta-2\eta^2}{2a_1^2}\;,\\
  v_\phi &= \xi_{\phi,3}^2\, p_3 + \xi_{\phi,4}^2 \,p_4
  = \frac{1+\eta-2\eta^2}{4\eta a_1^2}\;,
\end{aligned}
\end{equation}
which together gives $v_\eta+v_\phi=(1+3\eta-4\eta^3)/4\eta a_1^2$
saturating the Holevo bound~\eqref{n1HCR} as claimed.

At the boundary $\eta = (a_0^2-a_1^2)/2a_0^2$, the POVM outcome $\Pi_2=0$ while the remaining three
 reduce to a projective measurement on the eigenstate of
the SLD operator~\cite{Albarelli2019}
\begin{align}
\label{eq:3pi}  \Pi_1  &=
  \begin{pmatrix}
    1&0&0\\
    0&0&0\\
    0&0&0
  \end{pmatrix}\;,\qquad
  \begin{drcases}\Pi_{3}\\ \Pi_4\end{drcases}=\frac{1}{2}
  \begin{pmatrix}
    0&0&0\\
    0&1&\mp \I\\
    0&\pm \I&1
  \end{pmatrix}\;.  
\end{align}

In this case, the estimator coefficients are

\begin{align}
  \xi_{\eta,1}=-\frac{a_0^2+a_1^2
  \eta}{a_1^2}\;,\qquad \xi_{\eta,3}=\xi_{\eta,4}=1-\eta\;,\qquad \xi_{\phi,1}=0\;\qquad
  \text{and}\qquad\xi_{\phi,3}=-\xi_{\phi,4}=\frac{1}{2\sqrt{\eta}a_0a_1}\;.
\end{align}

This measurement scheme remains optimal even when
$\eta > (a_0^2-a_1^2)/2a_0^2$. Comparing the 4-outcome
POVM~\eqref{eq:4pi} to the 3-outcome POVM~\eqref{eq:3pi}, we see that
the role played by $\Pi_2$ is to obtain a better estimate of $\eta$,
but at the expense of a worse estimate of $\phi$. Whether this
trade-off improves the overall sum of the MSE depends on the exact
form of the probe and the value of $\eta$.  We note that the
estimators presented here depend on the unknown parameter
$\eta$. Although this would be an issue if we were interested in
global parameter estimation, for local estimation this is not an
issue, as we are only interested in estimating $\eta$ in the local
neighbourhood of some a priori known value, $\eta_0$.

%
\section{Dual solutions for the semidefinite program}
\label{apen:dual}
From the constructed POVM, we can arrive at a candidate for the
optimal $X$ and $\mathbb{L}$ matrices using (17) and
(14) from the main text which gives an upper bound to the primal solution. In
this appendix, we write down the dual problem and provide its solution
which gives a lower bound to the primal solution. One can easily check
that the lower and upper bounds coincide which implies that the
candidate solution is indeed an optimal solution for the
Nagaoka--Hayashi bound.

The dual problem is 
\begin{equation}
\begin{gathered}
\tilde{c}_\text{NH} = \max_{y  }\, \sum_k y_k\,c_k  \\
\text{subject to } \sum_k y_k  \bm{F}_k \leq \bm{F}_0\;,
\end{gathered}
\end{equation}
where the matrices $\bm{F}_k$ and constants $c_k$ implements the
constraints on the primal SDP as defined in appendix~\ref{app:sdp}.

\subsection{Qubit rotation estimation---dual solutions}
\label{apen:dual_ex1}
We first present the dual solution for the qubit rotation estimation problem. In order to write down the dual solutions, we need to choose a
representation for the set of basis matrices $\{B_j\}$ in
appendix~\ref{app:sdp}. We use the following 16 matrices:
\begin{equation}
\begin{aligned}
  B_1&=\frac{1}{2}\begin{pmatrix}1&0&0&0\\0&1&0&0\\0&0&1&0\\0&0&0&1\end{pmatrix},&
  B_2&=\frac{1}{\sqrt{2}}\begin{pmatrix}0&1&0&0\\1&0&0&0\\0&0&0&0\\0&0&0&0\end{pmatrix},&
  B_3&=\frac{1}{\sqrt{2}}\begin{pmatrix}0&0&1&0\\0&0&0&0\\1&0&0&0\\0&0&0&0\end{pmatrix},&
  B_4&=\frac{1}{\sqrt{2}}\begin{pmatrix}0&0&0&1\\0&0&0&0\\0&0&0&0\\1&0&0&0\end{pmatrix},\\
  B_5&=\frac{1}{\sqrt{2}}\begin{pmatrix}0&0&0&0\\0&0&1&0\\0&1&0&0\\0&0&0&0\end{pmatrix},&
  B_6&=\frac{1}{\sqrt{2}}\begin{pmatrix}0&0&0&0\\0&0&0&1\\0&0&0&0\\0&1&0&0\end{pmatrix},&
  B_7&=\frac{1}{\sqrt{2}}\begin{pmatrix}0&0&0&0\\0&0&0&0\\0&0&0&1\\0&0&1&0\end{pmatrix},&
  B_8&=\frac{1}{\sqrt{2}}\begin{pmatrix}0&-\I&0&0\\\I&0&0&0\\0&0&0&0\\0&0&0&0\end{pmatrix},\\
  B_9&=\frac{1}{\sqrt{2}}\begin{pmatrix}0&0&-\I&0\\0&0&0&0\\\I&0&0&0\\0&0&0&0\end{pmatrix},&
  B_{10}&=\frac{1}{\sqrt{2}}\begin{pmatrix}0&0&0&-\I\\0&0&0&0\\0&0&0&0\\\I&0&0&0\end{pmatrix},&
  B_{11}&=\frac{1}{\sqrt{2}}\begin{pmatrix}0&0&0&0\\0&0&-\I&0\\0&\I&0&0\\0&0&0&0\end{pmatrix},&
  B_{12}&=\frac{1}{\sqrt{2}}\begin{pmatrix}0&0&0&0\\0&0&0&-\I\\0&0&0&0\\0&\I&0&0\end{pmatrix},\\
  B_{13}&=\frac{1}{\sqrt{2}}\begin{pmatrix}0&0&0&0\\0&0&0&0\\0&0&0&-\I\\0&0&\I&0\end{pmatrix},&
  B_{14}&=\frac{1}{\sqrt{2}}\begin{pmatrix}1&0&0&0\\0&-1&0&0\\0&0&0&0\\0&0&0&0\end{pmatrix},&
  B_{15}&=\frac{1}{\sqrt{6}}\begin{pmatrix}1&0&0&0\\0&1&0&0\\0&0&-2&0\\0&0&0&0\end{pmatrix},&
  B_{16}&=\frac{1}{2\sqrt{3}}\begin{pmatrix}1&0&0&0\\0&1&0&0\\0&0&1&0\\0&0&0&-3\end{pmatrix}\,.
\end{aligned}
\end{equation}
One can then check that for estimating a single parameter, the following dual solution coincides with the
primal candidate:
\begin{equation}
\begin{gathered}
  y_{1,1}^{(2)}=1,\;
                 y^{(3)}_{1,2}=y^{(3)}_{1,7}=\frac{1}{\sqrt{8}},\;y^{(3)}_{1,3}=y^{(3)}_{1,6}=\frac{1-\epsilon}{\sqrt{8}},\,\\
  y^{(5)}_{1,1}=-\frac{1}{2},\,y^{(5)}_{1,4}=-\frac{1-\epsilon}{\sqrt{2}},\;y^{(5)}_{1,14}=-\frac{1}{\sqrt{8}},\;y^{(5)}_{1,15}=-\frac{1}{\sqrt{24}},\;y^{(5)}_{1,16}=\frac{1}{\sqrt{12}}\;,
\end{gathered}
\end{equation}
and all other $y_k$ zero. For estimating two parameters:
\begin{equation}
\begin{gathered}
  y_{1,1}^{(2)}=y^{(2)}_{2,2}=\frac{2}{2-\epsilon},\;y^{(3)}_{1,2}=y^{(3)}_{1,7}=y^{(3)}_{2,8}=y^{(3)}_{2,13}=\frac{1}{\sqrt{2}(2-\epsilon)},\;
  y^{(3)}_{1,3}=y^{(3)}_{1,6}=y^{(3)}_{2,9}=y^{(3)}_{2,12}=\frac{1-\epsilon}{\sqrt{2}(2-\epsilon)},\\
  y^{(4)}_{1,2,14}=-\frac{\epsilon}{\sqrt{8}},\;y^{(4)}_{1,2,15}=\epsilon\sqrt{\frac{3}{8}},\;
  y^{(5)}_{1}=-\frac{2}{2-\epsilon},\;y^{(5)}_{14}=-\frac{\sqrt{2}}{2-\epsilon},\;y^{(5)}_{15}=-\frac{\sqrt{2}}{\sqrt{3}(2-\epsilon)},\;y^{(5)}_{16}=\frac{2}{\sqrt{3}(2-\epsilon)},
\end{gathered}
\end{equation}
and all other  $y_k$ zero. For estimating three parameters:
\begin{equation}
\begin{gathered}
  y_{1,1}^{(2)}=y^{(2)}_{2,2}=\frac{2}{2-\epsilon},\;y^{(2)}_{3,3}=\frac{1}{(1-\epsilon)^2},\;
  y^{(3)}_{1,2}=y^{(3)}_{1,7}=y^{(3)}_{2,8}=y^{(3)}_{2,13}=\frac{1}{\sqrt{2}(2-\epsilon)},\\
  y^{(3)}_{1,3}=y^{(3)}_{1,6}=y^{(3)}_{2,9}=y^{(3)}_{2,12}=\frac{1-\epsilon}{\sqrt{2}(2-\epsilon)},\;y^{(3)}_{3,14}=-\frac{1}{\sqrt{8}},\;y^{(3)}_{3,15}=\frac{\sqrt{3}}{\sqrt{8}},\;
  y^{(4)}_{1,2,14}=-\frac{\epsilon}{\sqrt{8}},\;y^{(4)}_{1,2,15}=\epsilon\sqrt{\frac{3}{8}},\\
  y^{(5)}_{1}=-\frac{2}{2-\epsilon}-\frac{1}{2(1-\epsilon)^2},\;y^{(5)}_{1}=\frac{1}{\sqrt{2}(1-\epsilon)},\;y^{(5)}_{14}=-\frac{\sqrt{2}}{2-\epsilon}+\frac{\sqrt{2}}{4(1-\epsilon)^2},\;
  y^{(5)}_{15}=\frac{1}{\sqrt{3}}y^{(5)}_{14},\;y^{(5)}_{16}=-\sqrt{\frac{2}{3}}y^{(5)}_{14},
\end{gathered}
\end{equation}
and all other  $y_k$ zero.

\subsection{Phase and transmissivity estimation in
  interferometer---dual solutions}
\label{apen:dual_ex2}
We now write down the dual solution to the
Nagaoka--Hayashi bound for the second example, phase and transmissivity estimation in an interferometer, when $N=1$. To do this, we use the following 9 matrices as basis matrices:
\begin{equation}
\begin{aligned}
  B_1&=\frac{1}{\sqrt{3}}\begin{pmatrix}1&0&0\\0&1&0\\0&0&1\end{pmatrix},&
  B_2&=\frac{1}{\sqrt{2}}\begin{pmatrix}0&1&0\\1&0&0\\0&0&0\end{pmatrix},&
  B_3&=\frac{1}{\sqrt{2}}\begin{pmatrix}0&0&1\\0&0&0\\1&0&0\end{pmatrix},\\
  B_4&=\frac{1}{\sqrt{2}}\begin{pmatrix}0&0&0\\0&0&1\\0&1&0\end{pmatrix},&
  B_5&=\frac{1}{\sqrt{2}}\begin{pmatrix}0&-\I&0\\\I&0&0\\0&0&0\end{pmatrix},&
  B_6&=\frac{1}{\sqrt{2}}\begin{pmatrix}0&0&-\I\\0&0&0\\\I&0&0\end{pmatrix},\\
  B_7&=\frac{1}{\sqrt{2}}\begin{pmatrix}0&0&0\\0&0&-\I\\0&\I&0\end{pmatrix},&
  B_8&=\frac{1}{\sqrt{2}}\begin{pmatrix}1&0&0\\0&-1&0\\0&0&0\end{pmatrix},&
  B_9&=\frac{1}{\sqrt{6}}\begin{pmatrix}1&0&0\\0&1&0\\0&0&-2\end{pmatrix}\,.
\end{aligned}
\end{equation}
When $a_1<1/\sqrt{2}$ and $\eta<(a_0^2-a_1^2)/2a_0^2$, one solution to
the dual problem is:
\begin{equation}
\begin{gathered}
  y_{1,1}^{(2)}=\frac{(1-\eta)(1+2\eta)}{2a_1^2},\;
  y_{2,2}^{(2)}=\frac{(1-\eta)(1+2\eta)}{4\eta a_1^2},\\
  y^{(3)}_{1,7}=-y^{(3)}_{2,4}=\frac{a_0(1+\eta-2\eta^2)}{2\sqrt{2\eta} a_1},\;
  y^{(3)}_{2,8}=\frac{1+\eta-2\eta^2}{2\sqrt{2}},\;
  y^{(3)}_{2,9}=\sqrt\frac{3}{8}(1+\eta-2\eta^2),\\
  y_{1,2,1}^{(4)}=-\frac{1}{\sqrt{3}},\;
  y_{1,2,4}^{(4)}=-\sqrt{2\eta}a_0 a_1,\;
  y_{1,2,8}^{(4)}=\frac{a_0^2-(1-\eta)a_1^2}{\sqrt{2}},\;
  y_{1,2,9}^{(4)}=-\frac{1}{\sqrt{6}}+\sqrt{\frac{3}{2}}\eta a_1^2,\\
  y^{(5)}_{1}=-\frac{(1-\eta)(1+2\eta)^2}{4\sqrt{3}\eta a_1^2},\;
  y^{(5)}_{8}=-\frac{(1-\eta)(1+2\eta)^2}{4\sqrt{2}a_1^2},\;
  y^{(5)}_{9}=\frac{(1-\eta)(2-3\eta)(1+2\eta)^2}{4\sqrt{6}\eta a_1^2}\;,
\end{gathered}
\end{equation}
and all other $y_k$ zero. When the condition $a_1<1/\sqrt{2}$ and
$\eta<(a_0^2-a_1^2)/2a_0^2$ is not satisfied, one solution to the dual
problem is given by:
\begin{equation}
\begin{gathered}
  y_{1,1}^{(2)}=\frac{(1-\eta)(a_0^2+\eta a_1^2)}{a_1^2},\;
  y_{2,2}^{(2)}=\frac{a_0^2+\eta a_1^2}{4\eta a_0^2 a_1^2},\;
  y^{(3)}_{1,7}=\frac{(1-\eta)a_0}{\sqrt{2\eta}a_1}(a_0^2+\eta a_1^2),\;\\
  y^{(3)}_{2,4}=\frac{a_1^2 \eta -a_0^2 -8\eta(1-\eta)^2 a_1^2 a_0^4}{2\sqrt{2\eta}a_0 a_1},\;
  y^{(3)}_{2,8}=-\frac{1}{2\sqrt{2}}+\sqrt{2}a_0^2 (1-\eta)^2 (2a_0^2+\eta a_1^2),\;
  y^{(3)}_{2,9}=\sqrt{\frac{3}{8}}+\sqrt{6}\eta(1-\eta)^2 a_0^2 a_1^2, \\
  y_{1,2,1}^{(4)}=-\frac{2}{\sqrt{3}}(1-\eta)a_0^2,\;
  y_{1,2,4}^{(4)}=-2\sqrt{2\eta}(1-\eta)a_0^3 a_1,\;
  y_{1,2,8}^{(4)}=\sqrt{2}(1-\eta)a_0^2(a_0^2 -(1-\eta)a_1^2),\;\\
  y_{1,2,9}^{(4)}=-\sqrt{\frac{2}{3}}(1-\eta)a_0^2(1-3\eta a_1^2),\;
  y^{(5)}_{1}=-\frac{ (a_0^2+\eta a_1^2)(1+4\eta(1-\eta)a_0^2)}{4\sqrt{3}\eta a_0^2 a_1^2},\\
  y^{(5)}_{4}=\frac{1-4(1-\eta)^2a_0^4}{2\sqrt{2\eta}a_0 a_1},\;
  y^{(5)}_{8}=\frac{a_1^2 -4 a_0^2 (1-\eta)(a_0^2+a_0^2 a_1^2
    \eta+a_1^4\eta^2)}{4\sqrt{2}a_0^2 a_1^2},\;\\
  y^{(5)}_{9}=\frac{-\eta +(2+\eta+8\eta^2-20 \eta^3+12\eta^4)a_0^2+4\eta(1-\eta)^2(5-6\eta)a_0^4-12\eta(1-\eta)^2(2-\eta)a_0^6}{4\sqrt{6}\eta a_0^2 a_1^2}\;,
\end{gathered}
\end{equation}
and all other $y_k$ zero. One can check that these solutions coincide with the primal solution
in appendix~\ref{app:anal_povm}.

\bibliography{SDP_ref_ins}

\begin{thebibliography}{10}
\expandafter\ifx\csname url\endcsname\relax
  \def\url#1{\texttt{#1}}\fi
\expandafter\ifx\csname urlprefix\endcsname\relax\def\urlprefix{URL }\fi
\providecommand{\bibinfo}[2]{#2}
\providecommand{\eprint}[2][]{\url{#2}}

\bibitem{giovannetti2004quantum}
\bibinfo{author}{Giovannetti, V.}, \bibinfo{author}{Lloyd, S.} \&
  \bibinfo{author}{Maccone, L.}
\newblock \bibinfo{title}{Quantum-enhanced measurements: beating the standard
  quantum limit}.
\newblock \emph{\bibinfo{journal}{Science}} \textbf{\bibinfo{volume}{306}},
  \bibinfo{pages}{1330--1336} (\bibinfo{year}{2004}).

\bibitem{giovannetti2011advances}
\bibinfo{author}{Giovannetti, V.}, \bibinfo{author}{Lloyd, S.} \&
  \bibinfo{author}{Maccone, L.}
\newblock \bibinfo{title}{Advances in quantum metrology}.
\newblock \emph{\bibinfo{journal}{Nat. Photonics}}
  \textbf{\bibinfo{volume}{5}}, \bibinfo{pages}{222--229}
  (\bibinfo{year}{2011}).

\bibitem{aasi2013enhanced}
\bibinfo{author}{Aasi, J.} \emph{et~al.}
\newblock \bibinfo{title}{Enhanced sensitivity of the \text{LIGO} gravitational
  wave detector by using squeezed states of light}.
\newblock \emph{\bibinfo{journal}{Nat. Photonics}}
  \textbf{\bibinfo{volume}{7}}, \bibinfo{pages}{613--619}
  (\bibinfo{year}{2013}).

\bibitem{caves1981quantum}
\bibinfo{author}{Caves, C.~M.}
\newblock \bibinfo{title}{Quantum-mechanical noise in an interferometer}.
\newblock \emph{\bibinfo{journal}{Phys. Rev. D}} \textbf{\bibinfo{volume}{23}},
  \bibinfo{pages}{1693--1708} (\bibinfo{year}{1981}).

\bibitem{barnett2003ultimate}
\bibinfo{author}{Barnett, S.~M.}, \bibinfo{author}{Fabre, C.} \&
  \bibinfo{author}{Ma\^{i}tre, A.}
\newblock \bibinfo{title}{Ultimate quantum limits for resolution of beam
  displacements}.
\newblock \emph{\bibinfo{journal}{Eur. Phys. J. \textbf{D}}}
  \textbf{\bibinfo{volume}{22}}, \bibinfo{pages}{513--519}
  (\bibinfo{year}{2003}).

\bibitem{dorner2009optimal}
\bibinfo{author}{Dorner, U.} \emph{et~al.}
\newblock \bibinfo{title}{Optimal quantum phase estimation}.
\newblock \emph{\bibinfo{journal}{Phys. Rev. Lett}}
  \textbf{\bibinfo{volume}{102}}, \bibinfo{pages}{040403}
  (\bibinfo{year}{2009}).

\bibitem{kacprowicz2010experimental}
\bibinfo{author}{Kacprowicz, M.}, \bibinfo{author}{Demkowicz-Dobrza{\'n}ski,
  R.}, \bibinfo{author}{Wasilewski, W.}, \bibinfo{author}{Banaszek, K.} \&
  \bibinfo{author}{Walmsley, I.~A.}
\newblock \bibinfo{title}{Experimental quantum-enhanced estimation of a lossy
  phase shift}.
\newblock \emph{\bibinfo{journal}{Nat. Photonics}}
  \textbf{\bibinfo{volume}{4}}, \bibinfo{pages}{357--360}
  (\bibinfo{year}{2010}).

\bibitem{demkowicz2009quantum}
\bibinfo{author}{Demkowicz-Dobrza{\'n}ski, R.} \emph{et~al.}
\newblock \bibinfo{title}{Quantum phase estimation with lossy interferometers}.
\newblock \emph{\bibinfo{journal}{Phys. Rev. A}} \textbf{\bibinfo{volume}{80}},
  \bibinfo{pages}{013825} (\bibinfo{year}{2009}).

\bibitem{tsang2016quantum}
\bibinfo{author}{Tsang, M.}, \bibinfo{author}{Nair, R.} \& \bibinfo{author}{Lu,
  X.~M.}
\newblock \bibinfo{title}{Quantum theory of superresolution for two incoherent
  optical point sources}.
\newblock \emph{\bibinfo{journal}{Phys. Rev. X}} \textbf{\bibinfo{volume}{6}},
  \bibinfo{pages}{031033} (\bibinfo{year}{2016}).

\bibitem{tsang2019resolving}
\bibinfo{author}{Tsang, M.}
\newblock \bibinfo{title}{Resolving starlight: a quantum perspective}.
\newblock \emph{\bibinfo{journal}{Contemp. Phys.}}
  \textbf{\bibinfo{volume}{60}}, \bibinfo{pages}{279--298}
  (\bibinfo{year}{2019}).

\bibitem{yonezawa2012quantum}
\bibinfo{author}{Yonezawa, H.} \emph{et~al.}
\newblock \bibinfo{title}{Quantum-enhanced optical-phase tracking}.
\newblock \emph{\bibinfo{journal}{Science}} \textbf{\bibinfo{volume}{337}},
  \bibinfo{pages}{1514--1517} (\bibinfo{year}{2012}).

\bibitem{zhang2019quantum}
\bibinfo{author}{Zhang, L.} \emph{et~al.}
\newblock \bibinfo{title}{Quantum-limited fiber-optic phase tracking beyond
  $\pi$ range}.
\newblock \emph{\bibinfo{journal}{Opt. Express}} \textbf{\bibinfo{volume}{27}},
  \bibinfo{pages}{2327--2334} (\bibinfo{year}{2019}).

\bibitem{giovannetti2001quantum}
\bibinfo{author}{Giovannetti, V.}, \bibinfo{author}{Lloyd, S.} \&
  \bibinfo{author}{Maccone, L.}
\newblock \bibinfo{title}{Quantum-enhanced positioning and clock
  synchronization}.
\newblock \emph{\bibinfo{journal}{Nature}} \textbf{\bibinfo{volume}{412}},
  \bibinfo{pages}{417--419} (\bibinfo{year}{2001}).

\bibitem{lamine2008quantum}
\bibinfo{author}{Lamine, B.}, \bibinfo{author}{Fabre, C.} \&
  \bibinfo{author}{Treps, N.}
\newblock \bibinfo{title}{Quantum improvement of time transfer between remote
  clocks}.
\newblock \emph{\bibinfo{journal}{Phys. Rev. Lett.}}
  \textbf{\bibinfo{volume}{101}}, \bibinfo{pages}{123601}
  (\bibinfo{year}{2008}).

\bibitem{helstrom1968minimum}
\bibinfo{author}{Helstrom, C.~W.}
\newblock \bibinfo{title}{The minimum variance of estimates in quantum signal
  detection}.
\newblock \emph{\bibinfo{journal}{IEEE Trans. Inf. Theory}}
  \textbf{\bibinfo{volume}{14}}, \bibinfo{pages}{234--242}
  (\bibinfo{year}{1968}).

\bibitem{helstrom1967minimum}
\bibinfo{author}{Helstrom, C.~W.}
\newblock \bibinfo{title}{Minimum mean-squared error of estimates in quantum
  statistics}.
\newblock \emph{\bibinfo{journal}{Phys. Lett. A}}
  \textbf{\bibinfo{volume}{25}}, \bibinfo{pages}{101--102}
  (\bibinfo{year}{1967}).

\bibitem{belavkin1976generalized}
\bibinfo{author}{Belavkin, V.~P.}
\newblock \bibinfo{title}{Generalized uncertainty relations and efficient
  measurements in quantum systems}.
\newblock \emph{\bibinfo{journal}{Theor. Math. Phys.}}
  \textbf{\bibinfo{volume}{26}}, \bibinfo{pages}{213--222}
  (\bibinfo{year}{1976}).

\bibitem{robertson1929uncertainty}
\bibinfo{author}{Robertson, H.~P.}
\newblock \bibinfo{title}{The uncertainty principle}.
\newblock \emph{\bibinfo{journal}{Phys. Rev.}} \textbf{\bibinfo{volume}{34}},
  \bibinfo{pages}{163--164} (\bibinfo{year}{1929}).

\bibitem{cimini2019quantum}
\bibinfo{author}{Cimini, V.} \emph{et~al.}
\newblock \bibinfo{title}{Quantum sensing for dynamical tracking of chemical
  processes}.
\newblock \emph{\bibinfo{journal}{Phys. Rev. A}} \textbf{\bibinfo{volume}{99}},
  \bibinfo{pages}{053817} (\bibinfo{year}{2019}).

\bibitem{vrehavcek2017multiparameter}
\bibinfo{author}{{\v{R}}eha{\v{c}}ek, J.} \emph{et~al.}
\newblock \bibinfo{title}{Multiparameter quantum metrology of incoherent point
  sources: towards realistic superresolution}.
\newblock \emph{\bibinfo{journal}{Phys. Rev. A}} \textbf{\bibinfo{volume}{96}},
  \bibinfo{pages}{062107} (\bibinfo{year}{2017}).

\bibitem{ragy2016compatibility}
\bibinfo{author}{Ragy, S.}, \bibinfo{author}{Jarzyna, M.} \&
  \bibinfo{author}{Demkowicz-Dobrza{\'n}ski, R.}
\newblock \bibinfo{title}{Compatibility in multiparameter quantum metrology}.
\newblock \emph{\bibinfo{journal}{Phys. Rev. A}} \textbf{\bibinfo{volume}{94}},
  \bibinfo{pages}{052108} (\bibinfo{year}{2016}).

\bibitem{szczykulska2016multi}
\bibinfo{author}{Szczykulska, M.}, \bibinfo{author}{Baumgratz, T.} \&
  \bibinfo{author}{Datta, A.}
\newblock \bibinfo{title}{Multi-parameter quantum metrology}.
\newblock \emph{\bibinfo{journal}{Adv. Phys-X}} \textbf{\bibinfo{volume}{1}},
  \bibinfo{pages}{621--639} (\bibinfo{year}{2016}).

\bibitem{kull2020uncertainty}
\bibinfo{author}{Kull, I.}, \bibinfo{author}{Gu{\'e}rin, P.~A.} \&
  \bibinfo{author}{Verstraete, F.}
\newblock \bibinfo{title}{Uncertainty and trade-offs in quantum multiparameter
  estimation}.
\newblock \emph{\bibinfo{journal}{J. Phys. A}} \textbf{\bibinfo{volume}{53}},
  \bibinfo{pages}{244001} (\bibinfo{year}{2020}).

\bibitem{demkowicz2020multi}
\bibinfo{author}{Demkowicz-Dobrza{\'n}ski, R.}, \bibinfo{author}{G{\'o}recki,
  W.} \& \bibinfo{author}{Gu{\c{t}}{\u{a}}, M.}
\newblock \bibinfo{title}{Multi-parameter estimation beyond quantum {F}isher
  information}.
\newblock \emph{\bibinfo{journal}{J. Phys. A Math. Theor.}}
  \textbf{\bibinfo{volume}{53}}, \bibinfo{pages}{363001}
  (\bibinfo{year}{2020}).

\bibitem{suzuki2020quantum}
\bibinfo{author}{Suzuki, J.}, \bibinfo{author}{Yang, Y.} \&
  \bibinfo{author}{Hayashi, M.}
\newblock \bibinfo{title}{Quantum state estimation with nuisance parameters}.
\newblock \emph{\bibinfo{journal}{J. Phys. A Math. Theor.}}
  \textbf{\bibinfo{volume}{53}}, \bibinfo{pages}{453001}
  (\bibinfo{year}{2020}).

\bibitem{steinlechner2013quantum}
\bibinfo{author}{Steinlechner, S.} \emph{et~al.}
\newblock \bibinfo{title}{Quantum-dense metrology}.
\newblock \emph{\bibinfo{journal}{Nat. Photonics}}
  \textbf{\bibinfo{volume}{7}}, \bibinfo{pages}{626--630}
  (\bibinfo{year}{2013}).

\bibitem{hou2016achieving}
\bibinfo{author}{Hou, Z.}, \bibinfo{author}{Zhu, H.}, \bibinfo{author}{Xiang,
  G.~Y.}, \bibinfo{author}{Li, C.~F.} \& \bibinfo{author}{Guo, G.~C.}
\newblock \bibinfo{title}{Achieving quantum precision limit in adaptive qubit
  state tomography}.
\newblock \emph{\bibinfo{journal}{npj Quantum Inf.}}
  \textbf{\bibinfo{volume}{2}}, \bibinfo{pages}{16001} (\bibinfo{year}{2016}).

\bibitem{roccia2017entangling}
\bibinfo{author}{Roccia, E.} \emph{et~al.}
\newblock \bibinfo{title}{Entangling measurements for multiparameter estimation
  with two qubits}.
\newblock \emph{\bibinfo{journal}{Quantum Sci. Technol.}}
  \textbf{\bibinfo{volume}{3}}, \bibinfo{pages}{01LT01} (\bibinfo{year}{2018}).

\bibitem{liu2018loss}
\bibinfo{author}{Liu, Y.} \emph{et~al.}
\newblock \bibinfo{title}{Loss-tolerant quantum dense metrology with {SU}(1, 1)
  interferometer}.
\newblock \emph{\bibinfo{journal}{Opt. Express}} \textbf{\bibinfo{volume}{26}},
  \bibinfo{pages}{27705--27715} (\bibinfo{year}{2018}).

\bibitem{hou2018deterministic}
\bibinfo{author}{Hou, Z.} \emph{et~al.}
\newblock \bibinfo{title}{Deterministic realization of collective measurements
  via photonic quantum walks}.
\newblock \emph{\bibinfo{journal}{Nat. Commun.}} \textbf{\bibinfo{volume}{9}},
  \bibinfo{pages}{1414} (\bibinfo{year}{2018}).

\bibitem{humphreys2013quantum}
\bibinfo{author}{Humphreys, P.~C.}, \bibinfo{author}{Barbieri, M.},
  \bibinfo{author}{Datta, A.} \& \bibinfo{author}{Walmsley, I.~A.}
\newblock \bibinfo{title}{Quantum enhanced multiple phase estimation}.
\newblock \emph{\bibinfo{journal}{Phys. Rev. Lett.}}
  \textbf{\bibinfo{volume}{111}}, \bibinfo{pages}{070403}
  (\bibinfo{year}{2013}).

\bibitem{Genoni2013}
\bibinfo{author}{Genoni, M.~G.} \emph{et~al.}
\newblock \bibinfo{title}{Optimal estimation of joint parameters in phase
  space}.
\newblock \emph{\bibinfo{journal}{Phys. Rev. A}} \textbf{\bibinfo{volume}{87}},
  \bibinfo{pages}{012107} (\bibinfo{year}{2013}).

\bibitem{Crowley2014}
\bibinfo{author}{Crowley, P.}, \bibinfo{author}{Datta, A.},
  \bibinfo{author}{Barbieri, M.} \& \bibinfo{author}{Walmsley, I.~A.}
\newblock \bibinfo{title}{Tradeoff in simultaneous quantum-limited phase and
  loss estimation in interferometry}.
\newblock \emph{\bibinfo{journal}{Phys. Rev. A}} \textbf{\bibinfo{volume}{89}},
  \bibinfo{pages}{023845} (\bibinfo{year}{2014}).

\bibitem{gagatsos2016gaussian}
\bibinfo{author}{Gagatsos, C.~N.}, \bibinfo{author}{Branford, D.} \&
  \bibinfo{author}{Datta, A.}
\newblock \bibinfo{title}{Gaussian systems for quantum-enhanced multiple phase
  estimation}.
\newblock \emph{\bibinfo{journal}{Phys. Rev. A}} \textbf{\bibinfo{volume}{94}},
  \bibinfo{pages}{042342} (\bibinfo{year}{2016}).

\bibitem{baumgratz2016quantum}
\bibinfo{author}{Baumgratz, T.} \& \bibinfo{author}{Datta, A.}
\newblock \bibinfo{title}{Quantum enhanced estimation of a multidimensional
  field}.
\newblock \emph{\bibinfo{journal}{Phys. Rev. Lett.}}
  \textbf{\bibinfo{volume}{116}}, \bibinfo{pages}{030801}
  (\bibinfo{year}{2016}).

\bibitem{chrostowski2017super}
\bibinfo{author}{Chrostowski, A.}, \bibinfo{author}{Demkowicz-Dobrza{\'n}ski,
  R.}, \bibinfo{author}{Jarzyna, M.} \& \bibinfo{author}{Banaszek, K.}
\newblock \bibinfo{title}{On super-resolution imaging as a multiparameter
  estimation problem}.
\newblock \emph{\bibinfo{journal}{Int. J Quantum Inf.}}
  \textbf{\bibinfo{volume}{15}}, \bibinfo{pages}{1740005}
  (\bibinfo{year}{2017}).

\bibitem{pezze2017optimal}
\bibinfo{author}{Pezz{\`e}, L.} \emph{et~al.}
\newblock \bibinfo{title}{Optimal measurements for simultaneous quantum
  estimation of multiple phases}.
\newblock \emph{\bibinfo{journal}{Phys. Rev. Lett.}}
  \textbf{\bibinfo{volume}{119}}, \bibinfo{pages}{130504}
  (\bibinfo{year}{2017}).

\bibitem{suzuki2016explicit}
\bibinfo{author}{Suzuki, J.}
\newblock \bibinfo{title}{Explicit formula for the {H}olevo bound for
  two-parameter qubit-state estimation problem}.
\newblock \emph{\bibinfo{journal}{J Math. Phys.}}
  \textbf{\bibinfo{volume}{57}}, \bibinfo{pages}{042201}
  (\bibinfo{year}{2016}).

\bibitem{szczykulska2017reaching}
\bibinfo{author}{Szczykulska, M.}, \bibinfo{author}{Baumgratz, T.} \&
  \bibinfo{author}{Datta, A.}
\newblock \bibinfo{title}{Reaching for the quantum limits in the simultaneous
  estimation of phase and phase diffusion}.
\newblock \emph{\bibinfo{journal}{Quantum Sci. Technol.}}
  \textbf{\bibinfo{volume}{2}}, \bibinfo{pages}{044004} (\bibinfo{year}{2017}).

\bibitem{albarelli2020perspective}
\bibinfo{author}{Albarelli, F.}, \bibinfo{author}{Barbieri, M.},
  \bibinfo{author}{Genoni, M.~G.} \& \bibinfo{author}{Gianani, I.}
\newblock \bibinfo{title}{A perspective on multiparameter quantum metrology:
  from theoretical tools to applications in quantum imaging}.
\newblock \emph{\bibinfo{journal}{Phys. Lett. A}}
  \textbf{\bibinfo{volume}{384}}, \bibinfo{pages}{126311}
  (\bibinfo{year}{2020}).

\bibitem{assad2020accessible}
\bibinfo{author}{Assad, S.~M.} \emph{et~al.}
\newblock \bibinfo{title}{Accessible precisions for estimating two conjugate
  parameters using {G}aussian probes}.
\newblock \emph{\bibinfo{journal}{Phys. Rev. Res.}}
  \textbf{\bibinfo{volume}{2}}, \bibinfo{pages}{023182} (\bibinfo{year}{2020}).

\bibitem{tsang2019quantum}
\bibinfo{author}{Tsang, M.}, \bibinfo{author}{Albarelli, F.} \&
  \bibinfo{author}{Datta, A.}
\newblock \bibinfo{title}{Quantum semiparametric estimation}.
\newblock \emph{\bibinfo{journal}{Phys. Rev. X}} \textbf{\bibinfo{volume}{10}},
  \bibinfo{pages}{031023} (\bibinfo{year}{2020}).

\bibitem{carollo2019quantumness}
\bibinfo{author}{Carollo, A.}, \bibinfo{author}{Spagnolo, B.},
  \bibinfo{author}{Dubkov, A.~A.} \& \bibinfo{author}{Valenti, D.}
\newblock \bibinfo{title}{On quantumness in multi-parameter quantum
  estimation}.
\newblock \emph{\bibinfo{journal}{J. Stat. Mech.: Theory Exp.}}
  \textbf{\bibinfo{volume}{2019}}, \bibinfo{pages}{094010}
  (\bibinfo{year}{2019}).

\bibitem{sidhu2020geometric}
\bibinfo{author}{Sidhu, J.~S.} \& \bibinfo{author}{Kok, P.}
\newblock \bibinfo{title}{Geometric perspective on quantum parameter
  estimation}.
\newblock \emph{\bibinfo{journal}{AVS Quantum Sci.}}
  \textbf{\bibinfo{volume}{2}}, \bibinfo{pages}{014701} (\bibinfo{year}{2020}).

\bibitem{polino2020photonic}
\bibinfo{author}{Polino, E.}, \bibinfo{author}{Valeri, M.},
  \bibinfo{author}{Spagnolo, N.} \& \bibinfo{author}{Sciarrino, F.}
\newblock \bibinfo{title}{Photonic quantum metrology}.
\newblock \emph{\bibinfo{journal}{AVS Quantum Sci.}}
  \textbf{\bibinfo{volume}{2}}, \bibinfo{pages}{024703} (\bibinfo{year}{2020}).

\bibitem{suzuki2015parameter}
\bibinfo{author}{Suzuki, J.}
\newblock \bibinfo{title}{Parameter estimation of qubit states with unknown
  phase parameter}.
\newblock \emph{\bibinfo{journal}{Int. J. Quantum Inf.}}
  \textbf{\bibinfo{volume}{13}}, \bibinfo{pages}{1450044}
  (\bibinfo{year}{2015}).

\bibitem{bradshaw2018ultimate}
\bibinfo{author}{Bradshaw, M.}, \bibinfo{author}{Lam, P.~K.} \&
  \bibinfo{author}{Assad, S.~M.}
\newblock \bibinfo{title}{Ultimate precision of joint quadrature parameter
  estimation with a {G}aussian probe}.
\newblock \emph{\bibinfo{journal}{Phys. Rev. A}} \textbf{\bibinfo{volume}{97}},
  \bibinfo{pages}{012106} (\bibinfo{year}{2018}).

\bibitem{bradshaw2017tight}
\bibinfo{author}{Bradshaw, M.}, \bibinfo{author}{Assad, S.~M.} \&
  \bibinfo{author}{Lam, P.~K.}
\newblock \bibinfo{title}{A tight \text{C}ram{\'e}r--\text{R}ao bound for joint
  parameter estimation with a pure two-mode squeezed probe}.
\newblock \emph{\bibinfo{journal}{Phys. Lett. A}}
  \textbf{\bibinfo{volume}{381}}, \bibinfo{pages}{2598--2607}
  (\bibinfo{year}{2017}).

\bibitem{hayashi2005asymptotic}
\bibinfo{author}{Hayashi, M.}
\newblock \emph{\bibinfo{title}{Asymptotic Theory Of Quantum Statistical
  Inference: Selected Papers}} (\bibinfo{publisher}{World Scientific},
  \bibinfo{year}{2005}).

\bibitem{yuen1973}
\bibinfo{author}{Yuen, H.} \& \bibinfo{author}{Lax, M.}
\newblock \bibinfo{title}{Multiple-parameter quantum estimation and measurement
  of nonselfadjoint observables}.
\newblock \emph{\bibinfo{journal}{IEEE Trans. Inf. Theory}}
  \textbf{\bibinfo{volume}{19}}, \bibinfo{pages}{740--750}
  (\bibinfo{year}{1973}).

\bibitem{gill2005state}
\bibinfo{author}{Gill, R.~D.} \& \bibinfo{author}{Massar, S.}
\newblock \bibinfo{title}{State estimation for large ensembles}.
\newblock \emph{\bibinfo{journal}{Phys. Rev. A}} \textbf{\bibinfo{volume}{61}},
  \bibinfo{pages}{042312} (\bibinfo{year}{2000}).

\bibitem{holevo2011probabilistic}
\bibinfo{author}{Holevo, A.~S.}
\newblock \emph{\bibinfo{title}{Probabilistic and statistical aspects of
  quantum theory}}, vol.~\bibinfo{volume}{1} (\bibinfo{publisher}{Springer
  Science \& Business Media}, \bibinfo{year}{2011}).

\bibitem{Albarelli2019}
\bibinfo{author}{Albarelli, F.}, \bibinfo{author}{Friel, J.~F.} \&
  \bibinfo{author}{Datta, A.}
\newblock \bibinfo{title}{Evaluating the {H}olevo {C}ram{\'e}r-{R}ao bound for
  multiparameter quantum metrology}.
\newblock \emph{\bibinfo{journal}{Phys. Rev. Lett.}}
  \textbf{\bibinfo{volume}{123}}, \bibinfo{pages}{200503}
  (\bibinfo{year}{2019}).

\bibitem{sidhu2021tight}
\bibinfo{author}{Sidhu, J.~S.}, \bibinfo{author}{Ouyang, Y.},
  \bibinfo{author}{Campbell, E.~T.} \& \bibinfo{author}{Kok, P.}
\newblock \bibinfo{title}{Tight bounds on the simultaneous estimation of
  incompatible parameters}.
\newblock \emph{\bibinfo{journal}{Phys. Rev. X}} \textbf{\bibinfo{volume}{11}},
  \bibinfo{pages}{011028} (\bibinfo{year}{2021}).

\bibitem{Matsumoto2002}
\bibinfo{author}{Matsumoto, K.}
\newblock \bibinfo{title}{A new approach to the {C}ram{\'e}r-{Rao}-type bound
  of the pure-state model}.
\newblock \emph{\bibinfo{journal}{J. Phys. A: Math. Gen.}}
  \textbf{\bibinfo{volume}{35}}, \bibinfo{pages}{3111--3123}
  (\bibinfo{year}{2002}).

\bibitem{kahn2009local}
\bibinfo{author}{Kahn, J.} \& \bibinfo{author}{Gu{\c{t}}{\u{a}}, M.}
\newblock \bibinfo{title}{Local asymptotic normality for finite dimensional
  quantum systems}.
\newblock \emph{\bibinfo{journal}{Commun. Math. Phys.}}
  \textbf{\bibinfo{volume}{289}}, \bibinfo{pages}{597--652}
  (\bibinfo{year}{2009}).

\bibitem{yamagata2013quantum}
\bibinfo{author}{Yamagata, K.}, \bibinfo{author}{Fujiwara, A.} \&
  \bibinfo{author}{Gill, R.~D.}
\newblock \bibinfo{title}{Quantum local asymptotic normality based on a new
  quantum likelihood ratio}.
\newblock \emph{\bibinfo{journal}{Ann. Stat.}} \textbf{\bibinfo{volume}{41}},
  \bibinfo{pages}{2197--2217} (\bibinfo{year}{2013}).

\bibitem{yang2019attaining}
\bibinfo{author}{Yang, Y.}, \bibinfo{author}{Chiribella, G.} \&
  \bibinfo{author}{Hayashi, M.}
\newblock \bibinfo{title}{Attaining the ultimate precision limit in quantum
  state estimation}.
\newblock \emph{\bibinfo{journal}{Commun. Math. Phys.}}
  \textbf{\bibinfo{volume}{368}}, \bibinfo{pages}{223--293}
  (\bibinfo{year}{2019}).

\bibitem{nagaoka2005new}
\bibinfo{author}{Nagaoka, H.}
\newblock \bibinfo{title}{A new approach to {C}ram{\'e}r--{R}ao bounds for
  quantum state estimation}.
\newblock In \emph{\bibinfo{booktitle}{Asymptotic Theory Of Quantum Statistical
  Inference: Selected Papers}}, \bibinfo{pages}{100--112}
  (\bibinfo{publisher}{World Scientific}, \bibinfo{year}{2005}).
\newblock \bibinfo{note}{Originally published as IEICE Technical Report, 89,
  228, IT 89-42, 9-14, (1989)}.

\bibitem{nagaoka2005generalization}
\bibinfo{author}{Nagaoka, H.}
\newblock \bibinfo{title}{A generalization of the simultaneous diagonalization
  of {H}ermitian matrices and its relation to quantum estimation theory}.
\newblock In \emph{\bibinfo{booktitle}{Asymptotic Theory Of Quantum Statistical
  Inference: Selected Papers}}, \bibinfo{pages}{133--149}
  (\bibinfo{publisher}{World Scientific}, \bibinfo{year}{2005}).
\newblock \bibinfo{note}{Originally published as Trans. Jap. Soc. Indust. Appl.
  Math., 1, 43-56, (1991) in Japanese. Translated to English by Y.Tsuda.}

\bibitem{hayashi1999}
\bibinfo{author}{Hayashi, M.}
\newblock \bibinfo{title}{On simultaneous measurement of noncommutative
  observables}.
\newblock In \emph{\bibinfo{booktitle}{Development of infinite-dimensional
  non-commutative anaysis}}, \bibinfo{pages}{96--188}
  (\bibinfo{publisher}{Surikaisekikenkyusho (RIMS), Kyoto Univ., Kokyuroku No.
  1099, In Japanese}, \bibinfo{year}{1999}).

\bibitem{watanabe2011uncertainty}
\bibinfo{author}{Watanabe, Y.}, \bibinfo{author}{Sagawa, T.} \&
  \bibinfo{author}{Ueda, M.}
\newblock \bibinfo{title}{Uncertainty relation revisited from quantum
  estimation theory}.
\newblock \emph{\bibinfo{journal}{Phys. Rev. A}} \textbf{\bibinfo{volume}{84}},
  \bibinfo{pages}{042121} (\bibinfo{year}{2011}).

\bibitem{lofberg2004yalmip}
\bibinfo{author}{Lofberg, J.}
\newblock \bibinfo{title}{Yalmip: A toolbox for modeling and optimization in
  matlab}.
\newblock In \emph{\bibinfo{booktitle}{2004 IEEE international conference on
  robotics and automation (IEEE Cat. No. 04CH37508)}},
  \bibinfo{pages}{284--289} (\bibinfo{organization}{IEEE},
  \bibinfo{year}{2004}).

\bibitem{mosek}
\bibinfo{author}{ApS, M.}
\newblock \emph{\bibinfo{title}{The MOSEK optimization toolbox for MATLAB
  manual. Version 9.0.}}
\newblock \urlprefix\url{http://docs.mosek.com/9.0/toolbox/index.html}.
\newblock \bibinfo{note}{(2019)}.

\bibitem{boyd2004}
\bibinfo{author}{Boyd, S.} \& \bibinfo{author}{Vandenberghe, L.}
\newblock \emph{\bibinfo{title}{Convex optimization}}
  (\bibinfo{publisher}{Cambridge university press}, \bibinfo{year}{2004}).

\bibitem{huelga1997improvement}
\bibinfo{author}{Huelga, S.~F.} \emph{et~al.}
\newblock \bibinfo{title}{Improvement of frequency standards with quantum
  entanglement}.
\newblock \emph{\bibinfo{journal}{Phys. Rev. Lett.}}
  \textbf{\bibinfo{volume}{79}}, \bibinfo{pages}{3865--3868}
  (\bibinfo{year}{1997}).

\bibitem{myatt2000decoherence}
\bibinfo{author}{Myatt, C.~J.} \emph{et~al.}
\newblock \bibinfo{title}{Decoherence of quantum superpositions through
  coupling to engineered reservoirs}.
\newblock \emph{\bibinfo{journal}{Nature}} \textbf{\bibinfo{volume}{403}},
  \bibinfo{pages}{269--273} (\bibinfo{year}{2000}).

\bibitem{ma2011quantum}
\bibinfo{author}{Ma, J.}, \bibinfo{author}{Huang, Y.~x.},
  \bibinfo{author}{Wang, X.} \& \bibinfo{author}{Sun, C.~P.}
\newblock \bibinfo{title}{Quantum {F}isher information of the
  {G}reenberger-{H}orne-{Z}eilinger state in decoherence channels}.
\newblock \emph{\bibinfo{journal}{Phys. Rev. A}} \textbf{\bibinfo{volume}{84}},
  \bibinfo{pages}{022302} (\bibinfo{year}{2011}).

\bibitem{suzuki2019information}
\bibinfo{author}{Suzuki, J.}
\newblock \bibinfo{title}{Information geometrical characterization of quantum
  statistical models in quantum estimation theory}.
\newblock \emph{\bibinfo{journal}{Entropy}} \textbf{\bibinfo{volume}{21}},
  \bibinfo{pages}{703} (\bibinfo{year}{2019}).

\bibitem{Holland1993}
\bibinfo{author}{Holland, M.~J.} \& \bibinfo{author}{Burnett, K.}
\newblock \bibinfo{title}{Interferometric detection of optical phase shifts at
  the {H}eisenberg limit}.
\newblock \emph{\bibinfo{journal}{Phys. Rev. Lett.}}
  \textbf{\bibinfo{volume}{71}}, \bibinfo{pages}{1355--1358}
  (\bibinfo{year}{1993}).

\bibitem{suzuki2020nuisance}
\bibinfo{author}{Suzuki, J.}
\newblock \bibinfo{title}{Nuisance parameter problem in quantum estimation
  theory: {T}radeoff relation and qubit examples}.
\newblock \emph{\bibinfo{journal}{J. Phys. A Math. Theor.}}
  \textbf{\bibinfo{volume}{53}}, \bibinfo{pages}{264001}
  (\bibinfo{year}{2020}).

\bibitem{lin2015hiroshima}
\bibinfo{author}{Lin, M.} \& \bibinfo{author}{Wolkowicz, H.}
\newblock \bibinfo{title}{Hiroshima’s theorem and matrix norm inequalities}.
\newblock \emph{\bibinfo{journal}{Acta Sci. Math.(Szeged)}}
  \textbf{\bibinfo{volume}{81}}, \bibinfo{pages}{45--53}
  (\bibinfo{year}{2015}).

\bibitem{fujiwara1999estimation}
\bibinfo{author}{Fujiwara, A.} \& \bibinfo{author}{Nagaoka, H.}
\newblock \bibinfo{title}{An estimation theoretical characterization of
  coherent states}.
\newblock \emph{\bibinfo{journal}{J. Math. Phys.}}
  \textbf{\bibinfo{volume}{40}}, \bibinfo{pages}{4227--4239}
  (\bibinfo{year}{1999}).

\bibitem{hioe1981}
\bibinfo{author}{Hioe, F.~T.} \& \bibinfo{author}{Eberly, J.~H.}
\newblock \bibinfo{title}{{$N$}-level coherence vector and higher conservation
  laws in quantum optics and quantum mechanics}.
\newblock \emph{\bibinfo{journal}{Phys.Rev. Lett.}}
  \textbf{\bibinfo{volume}{47}}, \bibinfo{pages}{838--841}
  (\bibinfo{year}{1981}).

\bibitem{kimura2003}
\bibinfo{author}{Kimura, G.}
\newblock \bibinfo{title}{The {B}loch vector for {$N$}-level systems}.
\newblock \emph{\bibinfo{journal}{Phys. Lett. A}}
  \textbf{\bibinfo{volume}{314}}, \bibinfo{pages}{339--349}
  (\bibinfo{year}{2003}).

\bibitem{bertlmann2008}
\bibinfo{author}{Bertlmann, R.~A.} \& \bibinfo{author}{Krammer, P.}
\newblock \bibinfo{title}{{B}loch vectors for qudits}.
\newblock \emph{\bibinfo{journal}{J. Phys. A Math. Theor.}}
  \textbf{\bibinfo{volume}{41}}, \bibinfo{pages}{235303}
  (\bibinfo{year}{2008}).

\bibitem{vandenberghe1996semidefinite}
\bibinfo{author}{Vandenberghe, L.} \& \bibinfo{author}{Boyd, S.}
\newblock \bibinfo{title}{Semidefinite programming}.
\newblock \emph{\bibinfo{journal}{SIAM review}} \textbf{\bibinfo{volume}{38}},
  \bibinfo{pages}{49--95} (\bibinfo{year}{1996}).

\end{thebibliography}
\bibliographystyle{naturemag}

\end{document}